\documentclass{emulateapj}
\usepackage{epsfig}
\usepackage{rotating}
\usepackage{amsmath}
\usepackage{footnote}
\usepackage{courier}

\def \lowenvthresh {0.5}
\def \highenvthresh {3.0}
\def \apradius{2.5}
\def \apheight{35}

\begin{document}

\title{PRIMUS: Effect of Galaxy Environment on the Quiescent Fraction Evolution at \lowercase{\textit{z}}$<0.8$}
\author{
ChangHoon~Hahn\altaffilmark{1}, 
Michael R.~Blanton\altaffilmark{1}, 
John~Moustakas\altaffilmark{2},
Alison L.~Coil\altaffilmark{3}, 
Richard J.~Cool\altaffilmark{4}, 
Daniel J.~Eisenstein\altaffilmark{5},
Ramin A.~Skibba\altaffilmark{3}
Kenneth C.~Wong\altaffilmark{6}, 
Guangtun~Zhu\altaffilmark{7}
}
\altaffiltext{1}{Center for Cosmology and Particle Physics, Department of Physics, New York University, 4 Washington Place, New York, NY 10003; chh327@nyu.edu}
\altaffiltext{2}{Department of Physics and Astronomy, Siena College, 515 Loudon Road, Loudonville, NY 12211}
\altaffiltext{3}{Center for Astrophysics and Space Sciences, Department of Physics, University of California, 9500 Gilman Dr., La Jolla, CA 92093}
\altaffiltext{4}{MMT Observatory, University of Arizona, 1540 E Second Street, Tucson AZ 85721}
\altaffiltext{5}{Harvard-Smithsonian Center for Astrophysics, 60 Garden Street, Cambridge, MA 02138}
\altaffiltext{6}{Steward Observatory, University of Arizona, 933 North Cherry Avenue, Tucson, AZ 85721} 
\altaffiltext{7}{Hubble Fellow; Department of Physics and Astronomy, The Johns Hopkins University, 3400 North Charles Street, Baltimore, MD 21218} 
\begin{abstract}
We investigate the effects of galaxy environment on the evolution of
the quiescent fraction ($f_{\mathrm{Q}}$) from $z =0.8 $ to $ 0.0$ using
spectroscopic redshifts and multi-wavelength imaging data from the
PRIsm MUlti-object Survey (PRIMUS) and the Sloan Digitial Sky Survey
(SDSS). Our stellar mass limited galaxy sample consists of $\sim
14,000$ PRIMUS galaxies within $z = 0.2-0.8$ and $\sim 64,000$ SDSS
galaxies within $z = 0.05-0.12$. We classify the galaxies as quiescent
or star-forming based on an evolving specific star formation cut, and
as low or high density environments based on fixed cylindrical
aperture environment measurements on a volume-limited environment
defining population. For quiescent and
star-forming galaxies in low or high density environments, we examine
the evolution of their stellar mass function (SMF). Then using the
SMFs we compute $f_{\mathrm{Q}}(\mathcal{M}_{*})$ and quantify its
evolution within our redshift range. We find that the quiescent
fraction is higher at higher masses and in denser environments. The
quiescent fraction rises with cosmic time for all masses and
environments. At a fiducial mass of $10^{10.5}M_\odot$, from $z\sim
0.7$ to $0.1$, the quiescent fraction rises by $15\%$ at the
lowest environments and by $25\%$ at the highest environments we measure.
These results suggest that for a minority of galaxies their cessation
of star formation is due to external influences on
them. However, in the recent Universe a substantial fraction of the
galaxies that cease forming stars do so due to internal processes.
\end{abstract}
\section{Introduction}
Galaxies, in their detailed properties, carry the imprints of their surroundings, with a strong dependence of the quiescent fraction of galaxies on their local environment (e.g. \citealt{hubble36a, oemler74a, dressler80a, hermit96a, guzzo97a}; for a recent review see \citealt{blanton09a}).  The strength of this dependence is itself a strongly decreasing function of galaxy stellar mass; at the extreme, the lowest mass ($<10^{9}$ $M_\odot$) galaxies end their star formation only in dense regions, and never in isolation (\citealt{geha12a}). These effects also vary with redshift at least in the densest clusters, as observed in the changing fraction of late-type spirals relative to the field, found in studies of the morphology-density relation (\citealt{dressler84a, Fasano:2000aa, Smith:2005aa, desai07a}). Clearly understanding the properties of galaxies in the present-day universe requires a careful investigation of the role of environment, and how that role changes over time.


Nevertheless, the evolution of the role of environment is a relatively subtle effect and must be interpreted within the context of the evolving galaxy population. For instance, the most dramatic change in galaxy properties during the past eight billion years has been the remarkable decline in the star-formation rate of galaxies in the Universe (\citealt{hopkins06a}). This decline appears dominated by decreases in the rates of star-formation of individual galaxies (\citealt{Noeske:2007aa}). There is evidence that a large fraction of the decline is associated with strongly infrared-emitting starbursts (\citealt{bell05a, magnelli09a}). As \cite{cooper08a} and others have pointed out, because the environmental dependence of total star-formation rates at fixed redshift is relatively small, environmental effects are unlikely to cause the overall star-formation rate decline.

During this period, the major classes of galaxies that we observe today have already been firmly in place (\citealt{bundy06a, borch06a, taylor09a, Moustakas:2013aa}). Though not as dramatic as the history of galaxies prior to $z \sim 1$, detailed observations of the stellar mass function find significant evolution of the galaxy population with the decline in the number density of massive star-forming galaxies accompanied by an increase in the number density of quiescent galaxies (\citealt{Blanton:2006aa, bundy06a,  borch06a, Moustakas:2013aa}). \cite{Moustakas:2013aa}, for instance, find that since $z \sim 1.0$ the $\sim 50\%$ decline in the number density of massive star-forming galaxies ($\mathcal{M}_{*} > 10^{11} \mathcal{M}_{\odot}$) is complemented by the rise in number density of intermediate-mass quiescent galaxies ($\mathcal{M}_{*} \approx 10^{9.5} - 10^{10}\mathcal{M}_{\odot}$), by a factor of $ 2-3$, and massive quiescent galaxies ($\mathcal{M}_{*} > 10^{11} \mathcal{M}_{\odot}$ ), by $\sim 20\%$. On the color-magnitude diagram, this corresponds to the doubling of the red sequence over this period (\citealt{Bell:2004aa, borch06a, Faber:2007aa}). These changes in galaxy population are likely a result of physical processes that cause the cessation of star-formation in star-forming galaxies. 

Of the numerous mechanisms that have been proposed to explain this cessation, favored models suggest that internal processes such as supernovae or active galactic nuclei heat the gas within the galaxy, which consequently suppresses the cold gas supply used for star-formation (\citealt{Keres:2005aa, Croton:2006aa, Dekel:2008aa}). Other models propose that environment dependent external processes such as ram-pressure stripping (\citealt{Gunn:1972aa, Bekki:2009aa}), strangulation (\citealt{Larson:1980aa, Balogh:2000aa}), or harassment (\citealt{Moore:1998aa}) contribute to the cessation. 

Observations such as \cite{Weinmann:2006aa} and \cite{Peng:2010aa} credit some of these proposed internal processes for the cessation of star-formation, especially in massive galaxies. Meanwhile, observations of galaxy properties such as color and morphology correlating with environment suggest that environment may play a role in ceasing star-formation (\citealt{blanton09a} and references therein). However, it remains to be determined whether the environmental trends in galaxy properties reflect the direct effect of external environment on the galaxies' evolution (e.g. ram pressure, tidal forces, mergers) or reflect statistical differences in the histories of galaxies in different environments (e.g. an earlier formation time in dense regions).

In this paper we take the most straightforward investigation
by directly determining the star-forming properties of galaxies
as a function of environment, stellar mass and redshift in a single,
consistently analyzed data set. This analysis can reveal how galaxies
end their star formation over time, quantitatively establish the
contribution of environmental effects to the overall trends, and
reveal whether those trends happen equally in all environments.
However, such an analysis has not been done previously due to the lack
of sufficiently large samples. In this paper, we apply this approach
using the PRIism MUlti-object Survey (PRIMUS; \citealt{Coil:2011aa},
\citealt{Cool:2013aa}), the largest available redshift survey covering
the epochs between $0<z<1$.

In Section \ref{sec:sample} we present a brief description of the
PRIMUS and SDSS data, our mass complete sample construction, and
galaxy environment measurements. After dividing our galaxy sample into
subsamples of star-forming or quiescent and high or low density
environments, we compute and examine the evolution of the stellar mass
functions for our subsamples in Section \ref{sec:smf}. In
Section \ref{sec:qf_const}, we calculate the quiescent fraction,
analyze the evolution of the quiescent fraction, quantify the effects
of environment on the quiescent fraction evolution, and discuss the
implications of our quiescent fraction results on the cessation of
star-formation in galaxies. Finally in Section \ref{sec:summary} we
summarize our results.

Throughout the paper we assume a cosmology with $\Omega_{m} = 0.3,
\Omega_{\Lambda} = 0.7$, and $H_0 = 70 \: \mathrm{km} \; \mathrm{s}^{-1}
\mathrm{Mpc}^{-1}$. All magnitudes are AB-relative. 

\section{Sample Selection} \label{sec:sample}
We are interested in quantifying the effects of galaxy environment on
the evolution of the quiescent fraction over the redshift range $0 < z < 1$. For our analysis, we require a sample with sufficient depth and high quality spectroscopic redshift to probe the redshift range and to robustly measure galaxy environment. PRIMUS with its $\sim 120,000$ spectroscopic redshifts provides a large data set at intermediate redshifts for our analysis. In addition, we anchor our analysis with a low redshift sample derived from the Sloan Digital Sky Survey (\citealt{York:2000aa}). 

In Section \ref{sec:primus} and Section \ref{sec:sdss} we provide a
brief summary of the PRIMUS and SDSS data used for our sample
selection. In Section \ref{sec:target} we define our stellar mass complete galaxy sample.  
Then, in Section \ref{sec:sfq}, we classify the sample galaxies as quiescent or star-forming. 
We calculate the environment
using a volume-limited Environment Defining Population in Section
\ref{sec:environment}.  Finally, in Section \ref{sec:edgeeffect}, we
account for edge effects in the surveys.

\begin{figure*}
    \begin{center}
        \leavevmode
	\includegraphics[scale=0.475]{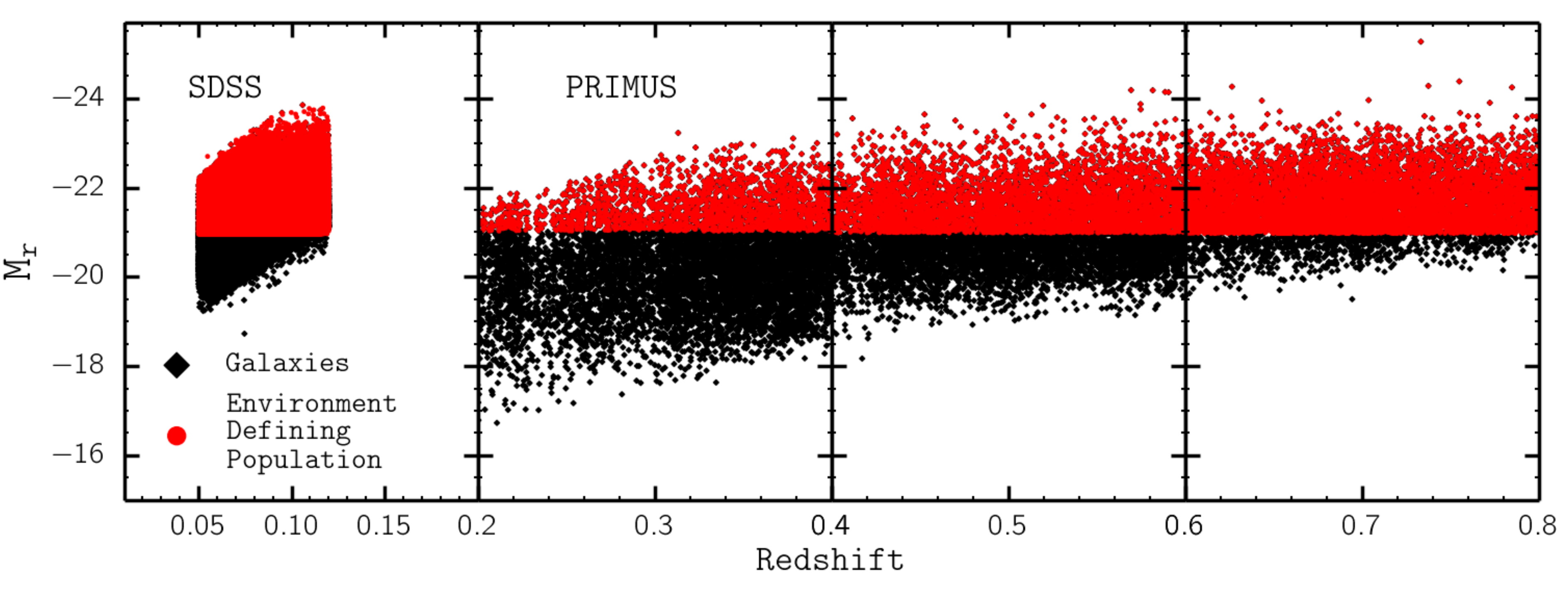}
        \caption{Absolute magnitude $M_{r}$ versus redshift for our mass complete galaxy sample (black squares) with the Environment Defining Population (red circles) plotted on top. Both samples are divided into redshift bins: $z = 0.05-0.12$, $0.2-0.4$, $0.4-0.6$, and $0.6-0.8$ (panels left to right). The lowest redshift bin ($z \approx 0.05-0.12$; leftmost panel) contain our galaxy sample and EDP selected from SDSS. The rest contain galaxies and EDP selected from PRIMUS. The redshift limits for the lowest redshift bin are empirically selected based on the bright and faint limits of SDSS galaxies. Stellar mass completeness limits, described in Section \ref{sec:target}, are imposed on the galaxy population. Meanwhile, $M_{r}$ limits are applied to the EDP such that the number density in each panel are equivalent (Section \ref{sec:environment}).} \label{fig:targetEDP}
    \end{center}
\end{figure*}
\subsection{PRIMUS} \label{sec:primus}
At intermediate redshifts we use multiwavelength imaging and
spectroscopic redshifts from PRIMUS, a faint galaxy survey with $\sim
120,000$ redshifts ($\sigma_z/(1+z) \approx 0.5 \%$) within the range
$z \approx 0-1.2$. The survey was conducted using the IMACS
spectrograph on the Magellan I Baade $6.5$-m telescope with a slitmask and low dispersion prism. For details on the PRIMUS observation methods such as survey design, targeting, and data summary, we refer readers to the survey papers (\citealt{Coil:2011aa, Cool:2013aa}). 

While the PRIMUS survey targeted seven distinct extragalactic deep fields for a total of $\sim 9 \; \mathrm{deg}^2$, we restrict our sample to five fields that have $GALEX$ and {\em Spitzer}/IRAC imaging for a total of $\sim 5.5 \; \mathrm{deg}^2$ (similar to the sample selection in \citealt{Moustakas:2013aa}). Four of these fields are a part of the {\em Spitzer} Wide-area Infrared Extragalactic Survey (SWIRE\footnote{http://swire.ipac.caltech.edu/swire/swire.html}): the European Large Area ISO Survey - South $1$ field (ELAIS-S1\footnote{http://dipastro.pd.astro.it/esis}), the Chandra Deep Field South SWIRE field (CDFS), and the XMM Large Scale Structure Survey field (XMM-LSS). The XMM-LSS consists of two separate but spatially adjacent fields: the Subaru/XMM-Newton DEEP Survey field (XMM-SXDSS\footnote{http://www.naoj.org/cience/SubaruProject/SDS}) and the Canadian-France-Hawaii Telescope Legacy Survey field (XMM-CFHTLS\footnote{http://www.cfht.hawaii.edu/Science/CFHLS}). Our fifth and final field is the Cosmic Evolution Survey (COSMOS\footnote{http://cosmos.astro.caltech.edu}) field. For all of our fields we have near-UV (NUV) and far-UV (FUV) photometry from the {\em GALEX} Deep Imaging Survey (DIS; \citealt{Martin:2005aa, Morrissey:2005aa}) as well as ground-based optical and {\em Spitzer}/IRAC mid-infrared photometric catalogs. \cite{Moustakas:2013aa} provides detailed descriptions of integrated flux calculations in the photometric bands for each of our fields. Furthermore, we derive the {\em K-corrections} from the photometry using \texttt{K-correct} (v4.2; \citealt{Blanton:2007aa}). 

Finally, using the spectroscopic redshift and broad wavelength photometry we apply \texttt{iSEDfit}, a Bayesian SED modeling code, to calculate stellar masses and star formation rates (SFRs) for our sample galaxies (\citealt{Moustakas:2013aa}). \texttt{iSEDfit} uses the redshift and the observed photometry of the galaxies to determine the statistical likelihood of a large ensemble of generated model SEDs. The model SEDs are generated using Flexible Stellar Population Synthesis (FSPS) models (\citealt{Conroy:2010aa}) based on the \cite{Chabrier:2003aa} IMF, along with other prior parameters discussed in Section 4.1 and Appendix A of \cite{Moustakas:2013aa}. For the observed photometry, we use the {\em GALEX} FUV and NUV, the two shortest IRAC bands at $3.6$ and $4.5 \mu \mathrm{m}$ (the two longer-wavelength IRAC channels are excluded because \texttt{iSEDfit} does not model hot dust or polycyclic aromatic hydrocarbons emission lines), and the optical bands. 

\subsection{SDSS-GALEX} \label{sec:sdss}
At low redshifts, we use spectroscopic redshifts and $ugriz$ photometry from the SDSS Data Release 7 (DR7; \citealt{Abazajian:2009aa}). More specifically we select galaxies from the New York University Value-Added Galaxy Catalog (hereafter VAGC) that satisfy the main sample criterion and have galaxy extinction corrected Petrosian magnitudes $14.5 < r < 17.6$ and spectroscopic redshifts within $0.01 < z < 0.2$ (\citealt{Blanton:2005aa}). We further restrict the VAGC sample to only galaxies with medium depth observations with total exposure time greater than $1 \; \mathrm{ks}$ from {\em GALEX} Release 6. This leaves $167,727$ galaxies. 

Next, we use the MAST/CasJobs\footnote{http://galex.stsci.edu/casjobs} interface and a $4''$ diameter search radius, to obtain the NUV and FUV photometry for the SDSS-{\em GALEX} galaxies. For optical photometry, we use the $ugriz$ bands from the SDSS \texttt{model} magnitudes scaled to the $r$-band \texttt{cmodel} magnitude. These photometric bands are then supplemented with integrated $JHK_s$ magnitudes from the 2MASS Extended Source Catalog (XSC; \citealt{Jarrett:2000aa}) and with photometry at $3.4$ and $4.6 \mu \mathrm{m}$ from the WISE All-Sky Data Release\footnote{http://wise2.ipac.caltech.edu/docs/release/allsky}. Further details regarding the SDSS-{\em GALEX} sample photometry can be found in Section 2.4 of \cite{Moustakas:2013aa}. As previously done on the PRIMUS data in Section \ref{sec:primus}, we use \texttt{iSEDfit} to obtain the stellar masses and star formation rates for the SDSS-{\em GALEX} sample. 

The SDSS-{\em GALEX} data discussed above is derived from the NYU-VAGC
based on SDSS Data Release 7, using the standard SDSS photometric
measurements. Several investigators have found that the background
subtraction techniques used in the standard photometric catalogs
introduce a size dependent bias in the galaxy fluxes and consequently
stellar masses (\citealt{West:2005aa, Blanton:2005ab, Lauer:2007aa, Bernardi:2007aa,
  Hyde:2009aa, West:2010aa}).

In order to quantify the effects of these photometric underestimations
in our analysis, we tried replacing our SDSS fluxes in the $ugriz$
band with $ugriz$ fluxes from the NASA-Sloan Atlas (NSA) catalog,
which incorporate the improved background subtraction presented in
\cite{Blanton:2011aa} and uses single-Seric fit fluxes rather than the
standard SDSS \texttt{cmodel} fluxes. Using the ratio of the
luminosity derived from the improved photometry over the luminosity
derived from the standard NYU-VAGC photometry, we apply a preliminary
correction to the stellar mass values obtained from \texttt{iSEDfit}
assuming a consistent mass-to-light ratio. This mass correction leads
to a significant increase in the stellar mass function for
$\mathcal{M} > 10^{11} \mathcal{M}_{\odot}$; however, the effect of
the mass correction was negligible for the quiescent fraction
evolution results. As a result, for the results presented here we use
the standard SDSS fluxes and we do not discuss the issues with
photometric measurements any further in this paper. We note that a
thorough investigation of these issues to understand their effect on
the stellar mass function requires a reanalysis of both the SDSS
photometry and the deeper photometry used for PRIMUS targeting.

\begin{table*} 
  \caption{Galaxy Subsamples}
  \label{tab:subsample}
  \begin{center}
    \leavevmode
    \begin{tabular}{ccccccc} \hline \hline              
     &\multicolumn{1}{c}{$n_{\mathrm{env}}$}        & \multicolumn{2}{c}{$N_{\mathrm{gal}}$}  & \multicolumn{2}{c}{$\mathcal{M}_{\mathrm{lim}}$} & $M_{\mathrm{r}, lim}$ \\ 
    & & Quiescent & Star-Forming & Quiescent & Star-Forming &  \\ \hline 
$0.05 < z < 0.12$ & $n_{\mathrm{env}} < \lowenvthresh $ & 6533 & 7508 & $10^{10.2} \mathcal{M}_{\odot}$ & $10^{10.2} \mathcal{M}_{\odot}$ & -20.95 \\
               & $n_{\mathrm{env}} > \highenvthresh $ &14673 & 9717 &                          \\ 
                              & all          &$33553$                       & $29864$                          \\ \hline
$0.2 < z < 0.4$      &$n_{\mathrm{env}} < \lowenvthresh $           &363                    &1231 & $10^{9.8} \mathcal{M}_{\odot}$ & $10^{9.8} \mathcal{M}_{\odot}$ &-21.03 \\
               &$n_{\mathrm{env}} > \highenvthresh $            &379                    &756                           \\
               & all                & $1086$                      & $2879$                          \\ \hline
$0.4 < z < 0.6$      &$n_{\mathrm{env}} < \lowenvthresh $           &536                       &1498 & $10^{10.3} \mathcal{M}_{\odot}$ & $10^{10.3} \mathcal{M}_{\odot}$ & -20.98 \\
               &$n_{\mathrm{env}} > \highenvthresh $            &490                       &854                           \\
               & all               & $1560$                      & $3577$                          \\ \hline
$0.6 < z < 0.8$      &$n_{\mathrm{env}} < \lowenvthresh $           &567                       &1254  & $10^{10.7} \mathcal{M}_{\odot}$ & $10^{10.6} \mathcal{M}_{\odot}$ & -20.97 \\
               &$n_{\mathrm{env}} > \highenvthresh $            &498                       &671                           \\
               & all              & $1668$                      & $2964$                          \\ \hline
Total &      & \multicolumn{2}{c}{77151} & \\ \hline
  \multicolumn{4}{l}{}                                             \\       
    \end{tabular} \par
    \end{center}
    {\bf Notes}: Number of galaxies ($N_{\mathrm{gal}}$) in the mass complete subsamples within the edges of the survey (Section \ref{sec:sample}). The subsamples are classified based on environment ($n_{\mathrm{env}}$) and star formation rate (star-forming or quiescent). The lowest redshift bin is derived from SDSS; the rest are from PRIMUS. We also list the stellar mass completeness limit, $\mathcal{M}_{\mathrm{lim}}$, for our sample along with the $r$-band absolute magnitude limits, $M_{\mathrm{r}, lim}$, for the Environment Defining Population. 
    \bigskip
\end{table*}

\subsection{Stellar Mass Complete Galaxy Sample} \label{sec:target} 
From the low redshift SDSS-{\em GALEX} and intermediate redshift
PRIMUS data we define our mass complete galaxy
sample. We begin by imposing the parent sample selection criteria from
\cite{Moustakas:2013aa}. More specifically, we take the statistically
complete {\em primary} sample from the PRIMUS data
(\citealt{Coil:2011aa}) and impose magnitude limits on optical
selection bands as specified in \cite{Moustakas:2013aa} Table 1. These
limits are in different optical selection bands and have distinct
values for the five PRIMUS target fields. We then exclude stars and
broad-line AGN to only select objects spectroscopically classified as
galaxies, with high-quality spectroscopic redshifts ($Q \geq
3$). Lastly, we impose a redshift range of $ 0.2 < z < 0.8$ for the
PRIMUS galaxy sample, where $ z > 0.2$ is selected due to limitations
from sample variance and $ z < 0.8$ is selected due to the lack of
sufficient statistics in subsamples defined below.

For the PRIMUS objects that meet the above criteria, we assign statistical weights (described in \citealt{Coil:2011aa} and \citealt{Cool:2013aa}) in order to correct for targeting incompleteness and redshift failures. The statistical weight, $w_i$, for each galaxy is given by
\begin{equation}
w_{i} = (f_{\mathrm{target}} \times f_{\mathrm{collision}} \times f_{\mathrm{success}})^{-1},
\end{equation}
as in Equation (1) in \cite{Moustakas:2013aa}. 

Since we are ultimately interested in a mass complete galaxy sample to
derive SMFs and QFs, next we impose stellar mass completeness limits
to our galaxy sample.
Stellar mass completeness limits for a magnitude-limited survey such as PRIMUS are functions of redshift, the apparent magnitude limit of the survey, and the typical stellar mass-to-light ratio of galaxies near the flux limit. We use the same procedure as \cite{Moustakas:2013aa}, which follows \cite{Pozzetti:2010aa}, to empircally determine the stellar mass completeness limits. For each of the target galaxies we compute $\mathcal{M}_{\mathrm{lim}}$ using $\log \; \mathcal{M}_{\mathrm{lim}} = \log \; \mathcal{M} + 0.4\;(m - m_{\mathrm{lim}})$, where $\mathcal{M}$ is the stellar mass of the galaxy in $\mathcal{M_{\odot}}$, $\mathcal{M}_{\mathrm{lim}}$ is the stellar mass of each galaxy if its magnitude was equal to the survey magnitude limit, $m$ is the observed apparent magnitude in the selection band, and $m_{\mathrm{lim}}$ is the magnitude limit for our five fields. We construct a cumulative distribution of $\mathcal{M}_{\mathrm{lim}}$ for the $15\%$ faintest galaxies in $\Delta z=0.04$ bins. In each of these redshift bins, we calculate the minimum stellar mass that includes $95 \%$ of the galaxies. Separately for quiescent and star-forming galaxies, we fit quadratic polynomials to the minimum stellar masses versus redshift (star-forming or quiescent classification is described in the following section). Finally, we use the polynomials to obtain the minimum stellar masses at the center of redshift bins, $0.2-0.4$, $0.4-0.6$, and $0.6-0.8$, which are then used as PRIMUS stellar mass completeness limits.

For the low redshift portion of our galaxy sample, we start by limiting the SDSS-{\em GALEX} data to objects within $0.05 < z < 0.12$, a redshift range later imposed on the volume-limited Environment Defining Population (Section \ref{sec:environment}). To account for the targeting incompleteness of the SDSS-{\em GALEX} sample, we use the statistical weight estimates provided by the NYU-VAGC catalog. Furthermore, we determine a uniform stellar mass completeness limit of $10^{10.2} \mathcal{M}_{\odot}$ above the stellar mass-to-light ratio completeness limit of the SDSS-{\em GALEX} data within the imposed redshift limits (\citealt{Blanton:2005ab, Baldry:2008aa, Moustakas:2013aa}). We then apply this mass limit in order to obtain our mass-complete galaxy sample at low redshift. 

We now have a stellar mass complete sample derived from SDSS-{\em
  GALEX} and PRIMUS data. Since our sample is derived from two
different surveys, we account for the disparity in the redshift
uncertainty. While PRIMUS provides a large number of redshifts out to
$z = 1$, due to its use of a low dispersion prism, the redshift
uncertainties are significantly larger ($\sigma_{z}/(1+z) \approx 0.5
\%$) than the uncertainties of the SDSS-{\em GALEX} redshifts. In order to have
comparable environment measures throughout our redshift range, we
apply PRIMUS redshift uncertainties to our galaxy sample selected from
SDSS-{\em GALEX}. For each SDSS-{\em GALEX} galaxy, we adjust its
redshift by randomly sampling a Gaussian distribution with standard
deviation $\sigma = 0.005 (1+z_{\mathrm{SDSS}-GALEX})$, where
$z_{\mathrm{SDSS}-GALEX}$ is the redshift of the galaxy.

\subsection{Classifying Quiescent and Star-Forming Galaxies} \label{sec:sfq}
We now classify our mass complete galaxy sample into quiescent or star-forming using an evolving cut based on specific star-formation rate utilized in \cite{Moustakas:2013aa} Section 3.2. This classification method uses the star-forming (SF) sequence, which is the correlation between star-formation rate (SFR) and stellar mass in star-forming galaxies observed at least until $z \sim 2$ (\citealt{Noeske:2007aa, Williams:2009aa, Karim:2011aa}). The PRIMUS sample displays a well-defined SF sequence within the redshift range of our galaxy sample. Using the power-law slope for the SF sequence from \cite{Salim:2007aa} (SFR $\propto \mathcal{M}^{0.65}$) and the minimum of the quiescent/star-forming bimodality, determined empirically, we obtain the following equation to classify the target galaxies (Equation 2 in \citealt{Moustakas:2013aa}):
\begin{equation} \label{eq:qsfclass} 
\mathrm{log}(\mathrm{SFR}_{\mathrm{min}}) = -0.49 + 0.64 \mathrm{log}(\mathcal{M} - 10) +1.07(z-0.1), 
\end{equation} 
where $\mathcal{M}$ is the stellar mass of the galaxy. If the target galaxy SFR and stellar mass lie above Equation \ref{eq:qsfclass} we classify it as star-forming; if below, as quiescent (\citealt{Moustakas:2013aa} Figure 1.).
\begin{figure}
  \begin{center}
    \leavevmode
    \includegraphics[scale=0.48]{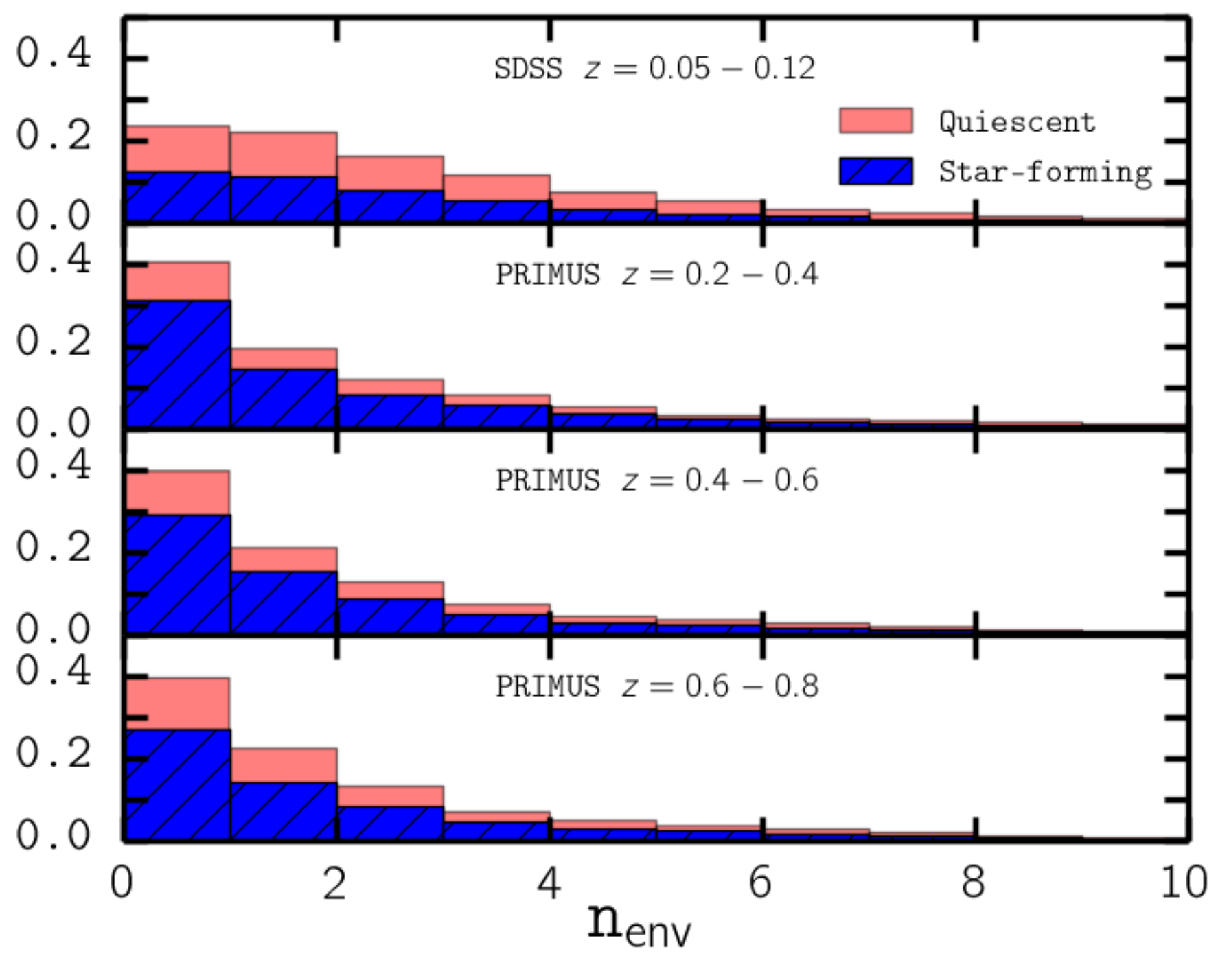}
    \caption{Normalized distribution of environment measurements ($n_{\mathrm{env}}$) for our mass complete galaxy sample within the survey edges. A fixed cylindrical aperture of $R_{\mathrm{ap}} =\apradius\;\mathrm{Mpc}$ and $H_{\mathrm{ap}} = \apheight\; \mathrm{Mpc}$ is used to measure environment. The star-forming galaxies contribution to the distribution is colored in blue and diagonally patterned. The contribution from quiescent galaxies is colored in red. Galaxies with $n_{\mathrm{env}} < 0.5$ are in low density environments and galaxies with $n_{\mathrm{env}} > 3.0$ are in high density environment. We note that the significant difference among the SDSS distribution and the PRIMUS distributions above is due to the different stellar mass completeness limits imposed on each redshift bin of our galaxy sample.}      \label{fig:envcount}
    \end{center}
\end{figure}

\subsection{Galaxy Environment} \label{sec:environment}
We define the environment of a galaxy as the number of neighboring Environment Defining Population galaxies (defined below) within a fixed aperture centered around it. We use fixed aperture measurements in order to quantify galaxy environment with an aperture sufficiently large to encompass massive halos (\citealt{Muldrew:2012aa, Skibba:2013aa}).

For our aperture, we use a cylinder of dimensions: $R_{\mathrm{ap}} =
\apradius\;
\mathrm{Mpc}$ and $H_{\mathrm{ap}} = \apheight\; \mathrm{Mpc}$.  We use a
cylindrical aperture to account for the PRIMUS redshift
errors and redshift space distortions (i.e. ``Finger of God"
effect). As \cite{Cooper:2005aa} and \cite{Gallazzi:2009aa} find, 
$\pm 1000 \; \mathrm{km} \;\mathrm{s^{-1}}$ optimally reduces the effects of redshift space
distortions. The PRIMUS redshift uncertainty at $z \sim 0.7$
corresponds to $\sigma_z < 0.01$, so our choice of $\apheight\;
\mathrm{Mpc}$ for the aperture height accounts for both of these effects. Our choice of
cylinder radius was motivated by scale dependence analyses in the
literature (\citealt{Blanton:2006aa, Wilman:2010aa, Muldrew:2012aa}),
which suggest that galactic properties such as color and quiescent
fractions are most strongly dependent on scales $< 2$ Mpc, around the
host dark matter halo sizes.



Before we measure the environment for our galaxy sample, we first
construct a volume limited Environment Defining Population (EDP) with
absolute magnitude cut-offs ($M_{r}$) in redshift bins with $\Delta z
\sim 0.2$. The $M_{r}$ cut-offs for the EDP are selected such that the
cumulative number density over $M_{r}$ for all redshift bins are
equal.  We make this choice in order to construct an EDP that contains
similar galaxy populations through the redshift range (i.e. accounts
for the progenitor bias). In their analysis of this method,
\cite{Behroozi:2013aa} and \cite{Leja:2013aa} find that although it
does not precisely account for the scatter in mass accretion or
galaxy-galaxy mergers, it provides a reasonable means to compare
galaxy populations over a wide range of cosmic time.

In constructing the PRIMUS EDP we use the same PRIMUS data used to select our galaxy sample (described in Section \ref{sec:target}). We again restrict the PRIMUS galaxies to $0.2 < z < 0.8$ and divide them into bins of $\Delta z = 0.2$. Before we consider the cumulative number densities in the redshift bins, we first determine the $M_r$ limit for the highest redshift bin ($z = 0.6-0.8$) by examining the $M_{r}$ distribution with bin size $\Delta M_{r} = 0.25$ and select $M_{r,\mathrm{lim}}$ near the peak of the distribution where bins with $M_{r} > M_{r,\mathrm{lim}}$ have fewer galaxies than the bin at $M_{r, \mathrm{lim}}$. We conservatively choose $M_{r, \mathrm{lim}}(0.6 < z < 0.8)$ to be $M_{r} = -20.97$. Then for the lower redshift bins, we impose absolute magnitude limits ($M_{r,\mathrm{lim}}$) such that the cumulative number density, calculated with the galaxy statistical weights, of the bin ordered by $M_{r}$ is equal to the cumulative number density of the highest redshift bin with $M_{r, \mathrm{lim}}(0.6 < z < 0.8) = -20.97$. 

For the SDSS EDP, we do not use the SDSS-{\em GALEX} parent data,
which is limited to the combined angular selection window of the
VAGC and {\em GALEX} (Section \ref{sec:sdss}). Instead, since FUV, NUV
values are not necessary for the EDP, we extend the parent data of the
SDSS EDP to the entire NYU-VAGC, including galaxies outside of the
{\em GALEX} window function. Furthermore, we impose a redshift range
of $0.05-0.12$ on the SDSS EDP. This redshift range was determined to
account for the lack of faint galaxies at $z \sim 0.2$ and the lack of
bright galaxies at $z \sim 0.01$ in the VAGC. As with the PRIMUS
redshift bins, we determine the SDSS EDP $M_{r, \mathrm{lim}}$ by matching
the cumulative number density of the highest redshift bin. For
redshift bins $z = 0.05-0.12$, $0.2-0.4$, $0.4-0.6$, $0.6-0.8$ we get
$M_{r,\mathrm{lim}} = -20.95$, $-21.03$, $-20.98$ and $-20.97$,
respectively. These absolute magnitude limits are illustrated in
Figure \ref{fig:targetEDP}, where we present the absolute magnitude ($M_{r}$) versus redshift for the galaxy sample (black squares) ad the EDP (red circles). 
The left-most panel corresponds to the samples derived from the SDSS-{\em GALEX} data while the rest correspond to samples derived from the PRIMUS data divided in bins with $\Delta z \sim 0.2$. 
Figure \ref{fig:targetEDP} shows clear $M_r$ cutoffs in the
$M_{r}$ distribution versus redshift for the EDP on top
of our galaxy sample.

For our SDSS-{\em GALEX} galaxy sample, in Section \ref{sec:target}, we apply PRIMUS redshift errors in order to establish a consistent measurement of environment throughout our redshift range. We appropriately apply equivalent redshift adjustments for the SDSS EDP. For the SDSS EDP galaxies that are also contained within the SDSS-{\em GALEX} sample, we adjust the redshift by an identical amount. For the rest, we apply the same redshift adjustment procedure described in Section \ref{sec:target} in order to obtain PRIMUS level redshift uncertainties. 

Finally, we measure the environment for each galaxy in our galaxy
sample by counting the number of EDP galaxies, $n_{\mathrm{env}}$, with RA,
Dec, and $z$ within our cylindrical aperture centered around
it. $n_{\mathrm{env}}$ accounts for the statistical weights of the EDP
galaxies. 
For our galaxy sample, the expected $n_{\mathrm{env}}$ given the uniform number density in 
each of our EDP redshift bin and volume of our cylindrical aperture is $\langle n_{\mathrm{env}} \rangle = 1.3$. 
Once we obtain environment measurements for all the galaxies
in our galaxy sample, we classify galaxies with $n_{\mathrm{env}} < 0.5$
to be in ``low" environment densities and galaxies with $n_{\mathrm{env}} > 3$
to be in ``high" environment densities. The high environment cutoff was
selected in order to reduce contamination from galaxies in low
environment densities while maintaining sufficient statistics. In
Section \ref{sec:env_qf_evol} we will also explore higher density
cutoffs for $n_{\mathrm{env}}$.

The analysis we describe below uses a fixed cylindrical aperture with
dimensions $R_{\mathrm{ap}} = \apradius \; \mathrm{Mpc}$ and $H_{\mathrm{ap}} = \apheight
\;\mathrm{Mpc}$ to measure environment. The same analysis was
extended for varying aperture dimensions $R_{\mathrm{ap}} = 1.5, \: \apradius, \:3.0 \:
\mathrm{Mpc}$ and $H_{\mathrm{ap}} = \apheight, \; 70 \;\mathrm{Mpc}$ with adjusted environment classifications. 
The results obtained from using different apertures and
environment classifications are qualitatively consistent with the results presented below.  

\subsection{Edge Effects} \label{sec:edgeeffect}
One of the challenges in obtaining accurate galaxy environments using a fixed aperture method is accounting for the edges of the survey. For galaxies located near the edge of the survey, part of the fixed aperture encompassing it will lie outside the survey regions. In this scenario, $n_{env}$ only reflects the fraction of the environment within the survey geometry.

To account for these edge effects, we use a Monte Carlo method to impose edge cutoffs on our galaxy sample. First, using \texttt{ransack} from \cite{Swanson:2008aa}, we construct a random sample of  $N_{\mathrm{ransack}} = 1,000,000$ points with RA and Dec randomly selected within the window function of the EDP (SDSS EDP and PRIMUS EDP separately). We then compute the angular separation, $\theta_{i, \mathrm{ap}}$ that corresponds to $R_{\mathrm{ap}}$ (Section \ref{sec:environment}) at the redshift of each sample galaxy $i$. For each sample galaxy we count the number of \texttt{ransack} points within $\theta_{i, \mathrm{ap}}$ of the galaxy: $n_{i,\mathrm{ransack}}$. Afterwards, we compare $n_{i,\mathrm{ransack}}$ to the expected value computed from the angular area of the environment defining aperture and the EDP window function: 
\begin{equation} \label{eq:ransack}
\langle n_{\mathrm{ransack}}\rangle_{i} = \frac{N_{\mathrm{ransack}}}{A_{\mathrm{EDP}}}\times {\pi \theta_{i, \mathrm{ap}}^2} \times f_{\mathrm{thresh}}. 
\end{equation} 
$A_{\mathrm{EDP}}$ is the total angular area of the EDP window function and $f_{\mathrm{thresh}}$ is the fractional threshold for the edge effect cut-off. For $R_{\mathrm{ap}}= \apradius \;\mathrm{Mpc}$, we use $f_{\mathrm{thresh}} = 0.75$. If $n_{i, \mathrm{ransack}} > \langle n_{\mathrm{ransack}} \rangle_i$ then galaxy $i$ remains in our sample; otherwise, it is discarded. Once the edge effect cuts are applied, we are left with the final galaxy sample. For our SDSS-{\em GALEX} galaxy sample, $\sim 12 \%$ of galaxies are removed from the edge effect cuts. For our PRIMUS galaxy sample, $\sim 40 \%$ of galaxies are removed from the edge effect cuts. 

In Figure \ref{fig:envcount} we present the distribution of environment measurements ($n_{\mathrm{env}}$) for our final galaxy sample in redshift bins: $z = 0.05 - 0.12$, $0.2 - 0.4$, $0.4-0.6$, and $0.6-0.8$. The quiescent galaxy contributions are colored in red while the star-forming galaxy contributions are colored in blue and patterned. We classify galaxies with $n_{\mathrm{env}} < 0.5$ to be in low density environments and galaxies with $n_{\mathrm{env}} > 3.0$ to be in high density environments. 

Although we imposed PRIMUS redshift errors on our SDSS galaxies to consistently measure environment throughout our entire sample, we note a significant discrepancy between the $n_{\mathrm{env}}$ distributions of the SDSS and PRIMUS samples. For example, in each of the PRIMUS redshift bins, $\sim 40 \%$ of galaxies in the redshift bin are in low density environments and roughly $30 \%$ are in high density environments. In contrast, in the SDSS redshift bin, $\sim 20 \%$ of galaxies in the redshift bin are in low density environments and $\sim 35 \%$ are in high density environments. We remind the reader that this is mainly due to the varying stellar mass-completeness limits imposed on our galaxy sample for each redshift bins and does not affect our results. 
\begin{figure*}
  \begin{center}
    \leavevmode
    \epsfig{file=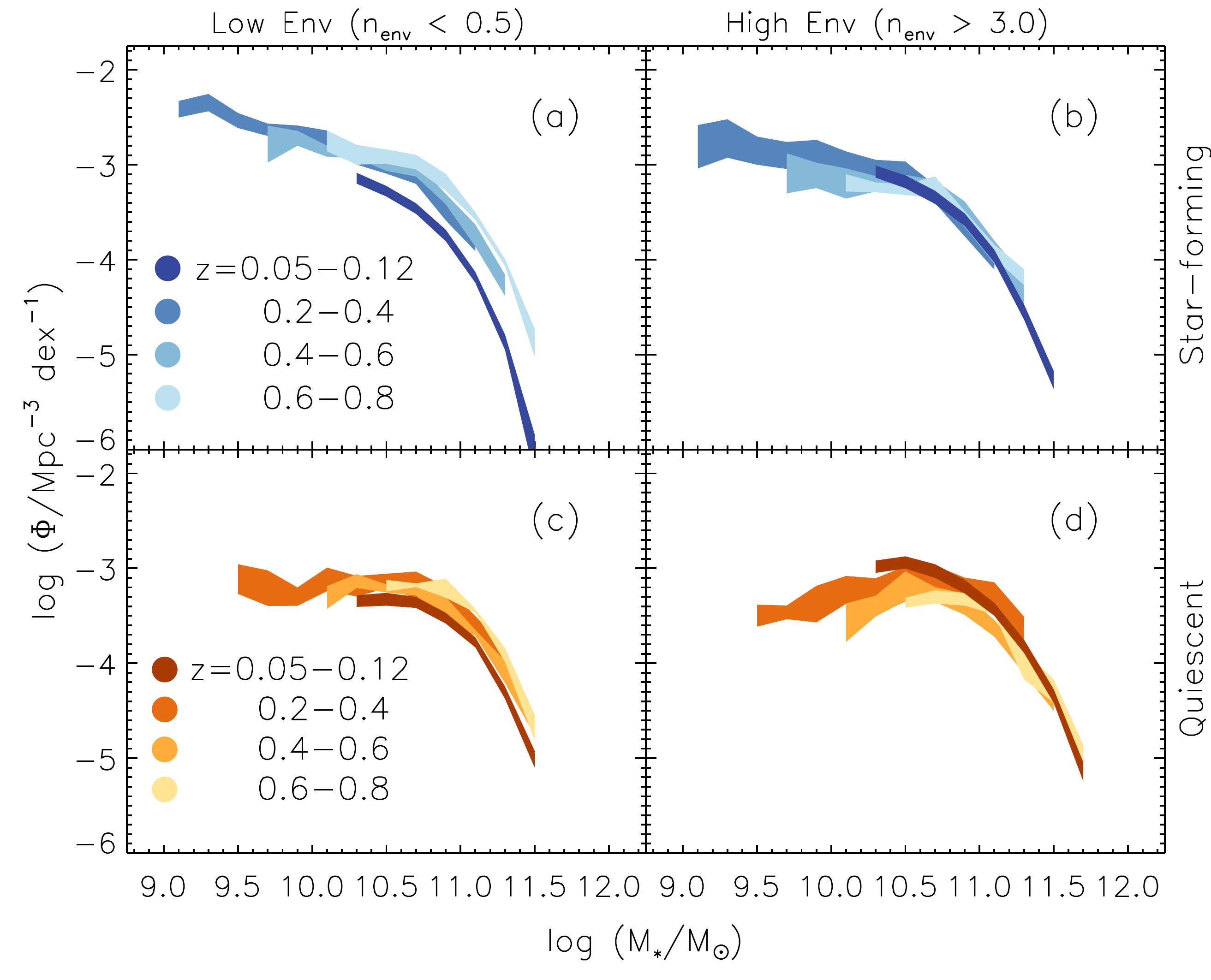, height=0.75\textwidth,angle=90}
     \caption{Evolution of stellar mass functions of star-forming (top) and quiescent (bottom) galaxies in 
low (left) and high (right) density environments throughout the redshift range
$z=0$--$0.8$. The environment of each galaxy  
was calculated using a cylindrical aperture size of $R=\apradius \: \mathrm{Mpc}$ and $H=\apheight \: \mathrm{Mpc}$ and  
classified as low environment when $n_{\mathrm{env}} < 0.5$ and as high environment when $n_{\mathrm{env}} > 3.0$. The SMFs use mass bins of 
width $\Delta \mathrm{log}(\mathcal{M}/\mathcal{M}_{\odot})=0.2$. In each panel we use shades of blue 
(star-forming) and orange (quiescent) to represent the SMF at different redshift, higher redshifts being
progressively lighter.}      \label{fig:smf}
    \end{center}
\end{figure*}
\section{Results: Stellar Mass Function} \label{sec:smf}
Our galaxy sample has so far been classified into quiescent or
star-forming and low or high density environments. We further divide
these subsamples into redshift bins: $0.05-0.12$, $0.2-0.4$,
$0.4-0.6$, and $0.6-0.8$ for a total of 16 subsamples. In Section \ref{sec:smfcalc}, we calculate the SMF for each of these subsamples. Then we examine the evolution of active and quiescent subsample SMFs in different environments in Section \ref{sec:smfevol}.  
\subsection{Stellar Mass Function Calculations} \label{sec:smfcalc} 
To calculate the SMFs we employ a non-parametric $1/{V_{\mathrm{max}}}$ estimator commonly used for galaxy luminosity functions and stellar mass functions in order to account for Malmquist bias, as done in \cite{Moustakas:2013aa} and discussed in the review \cite{Johnston:2011aa}. The differential SMF is given by the following equation:
\begin{equation} \label{eq:phi}
\Phi(\mathrm{log}\: \mathcal{M}) \Delta(\mathrm{log} \:\mathcal{M}) = \sum\limits_{i=1}^{N} \frac{w_i}{V_{\mathrm{max,avail},i}}. 
\end{equation}
$w_i$ is the statistical weight of galaxy $i$ and $\Phi(\mathrm{log}\:
\mathcal{M}) \Delta(\mathrm{log}\: \mathcal{M})$ is the number of galaxies
($N$) per unit volume within the stellar mass range $[\mathrm{log}
  \mathcal{M},\: \mathrm{log} \mathcal{M}+\Delta(\mathrm{log}\mathcal{M})]$. The equation above is the same as Equation 3 in \cite{Moustakas:2013aa} except that we use $V_{\mathrm{max,avail}}$ instead than $V_{\mathrm{max}}$, to account for the edge effects of the survey discussed in Section \ref{sec:edgeeffect}. 

$V_{\mathrm{max},i}$ is the maximum cosmological volume where it is
possible to observe galaxy $i$ given the apparent magnitude limits of
the survey. However in Section \ref{sec:edgeeffect} we remove galaxies
that lie on the survey edges from our sample. In doing so, we reduce the
maximum cosmological volume where a galaxy can be observed, thereby
reducing the fraction of $V_{\mathrm{max},i}$ that is actually available
in the sample. We introduce the term $V_{\mathrm{max,avail},i}$ to express
the maximum volume accounting for the survey edge effects.

To calculate $V_{\mathrm{max,avail},i}$, we use a similar Monte Carlo
method as the edge effect cutoffs in Section
\ref{sec:edgeeffect}. First, we generate a sample of points with
random RA, Dec within the window function of our galaxy sample
(SDSS-{\em GALEX} window function and the five PRIMUS fields) and
random $z$ within the redshift range. These points are not to be
confused with the \texttt{ransack} sample in Section
\ref{sec:edgeeffect}. We apply the edge effect cuts on these random
points as we did for our galaxy sample using the same method as in
Section \ref{sec:edgeeffect}. Within redshift bins of $\Delta z \sim
0.01$, we calculate the fraction of the random points that remain in
the bin after the edge effect cuts over the total number of random
points in the bin: $f_{\mathrm{edge}}$. We then apply this factor to
compute $V_{\mathrm{max,avail}} = V_{\mathrm{max}} \times f_{\mathrm{edge}}$. The
$V_{\mathrm{max}}$ values in the equation above are computed following the
method described in \cite{Moustakas:2013aa} Section 4.2 with the same
redshift-dependent $K$-correction from the observed SED and luminosity
evolution model.

To calculate the uncertainty of the SMFs from the sample variance, we use a standard jackknife technique (following \citealt{Moustakas:2013aa}). For the PRIMUS galaxies, we calculate SMFs after excluding one of the five target fields at a time. For the SDSS target galaxies we divide the field into a 12 $\times$ 9 rectangular RA and Dec grid and calculate the SMFs after excluding one grid at a time. From the calculated SMFs we calculate the uncertainty: 
\begin{equation}
\sigma^j = \sqrt{\frac{N-1}{N} \sum\limits_{k=1}^{M} (\Phi^j_k - \langle \Phi^j \rangle)^2}
\end{equation} 
$N$ in this equation is the number of jackknife SMFs in the stellar mass bin. $\langle \Phi^j \rangle$ is the mean number density of galaxies in each stellar mass bin for all of the jackknife $\Phi^j$s. 

\subsection{Evolution of the Stellar Mass Function in Different Environments} \label{sec:smfevol}
In Figure \ref{fig:smf}, we present the SMFs of the quiescent/star-forming (orange/blue, bottom/top panels) and high/low density environment (left/right panels) subsamples. The redshift evolution of the SMFs in each of these panels are indicated by a darker shade for lower redshift bins. The width of the SMFs represent the sample variance uncertainties derived in Section \ref{sec:smfcalc}.

While a detailed comparison of the SMFs in each panel for different
epochs is complicated by the different stellar mass completeness limits, we present some notable trends in each panel. In panel (a), star-forming galaxies in low density environments, we find a significant decrease in the high mass end of the SMF ($\mathcal{M} > 10^{10.75} \mathcal{M}_{\odot}$) over cosmic time. Meanwhile at lower masses ($\mathcal{M} < 10^{10.5} \mathcal{M}_{\odot}$), we observe no noticeable trend in the SMF. In panel (b), star-forming galaxies in high density environments, we do not observe any clear trends above the knee of the SMF ($\mathcal{M} \sim 10^{10.7} \mathcal{M}_{\odot}$) but an increase in SMF below the knee. For the quiescent population in low density environment, panel (c), we observe a potential decrease at higher masses ($\mathcal{M} > 10^{10.7} \mathcal{M}_{\odot}$). Lastly for the quiescent population in high density environments, panel (d), we find significant increase in $\Phi$ for lower masses but little trend at higher masses. 


\begin{figure*}
    \begin{center}
        \leavevmode
        \epsfig{file=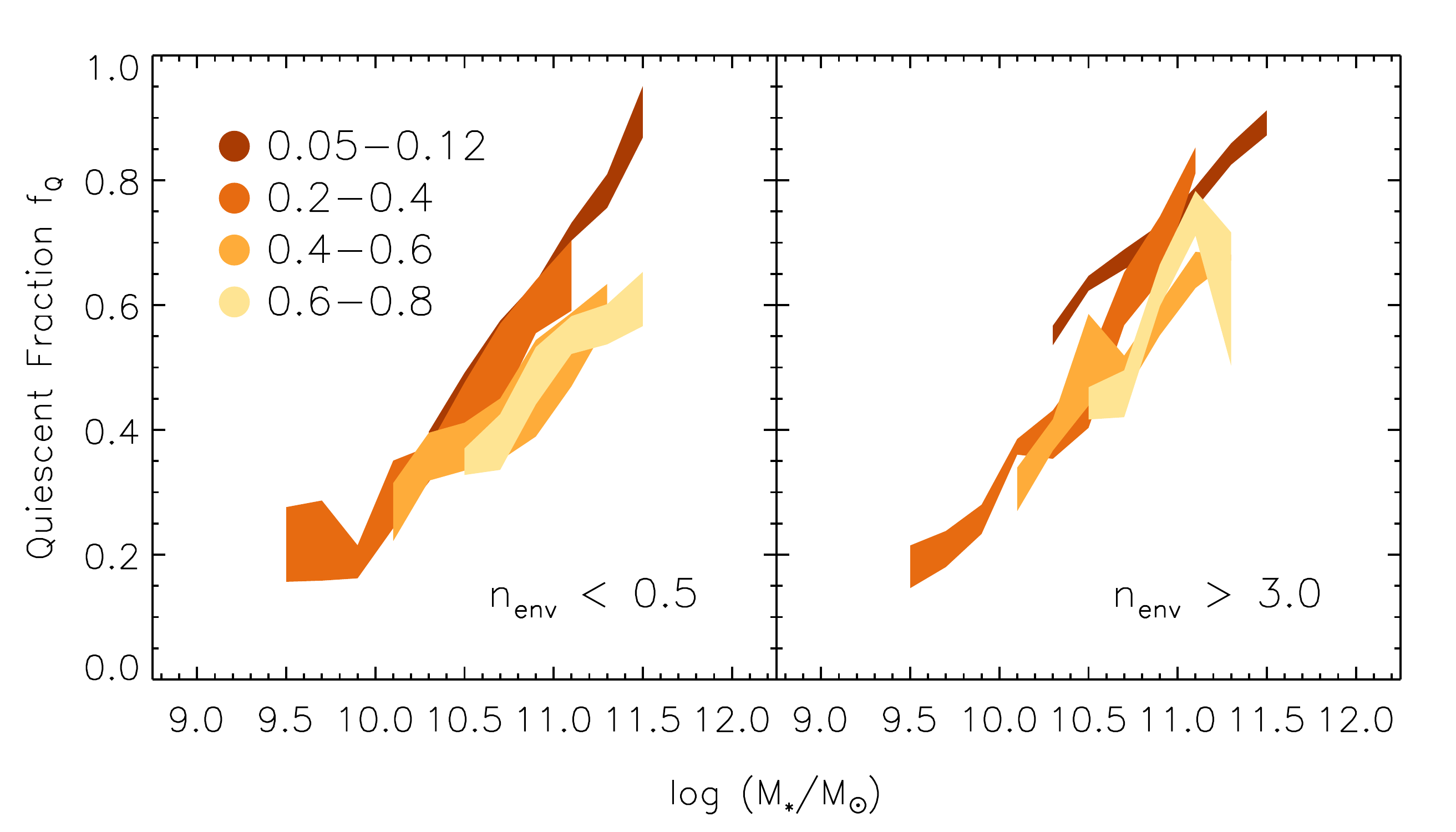, height=0.75\textwidth,angle=90}
        \caption{Evolution of the quiescent fraction $f_{\mathrm{Q}}$ for
          galaxies in low (left) and high (right) density
          environments for $z < 0.8$. $f_{\mathrm{Q}}$s were calculated
          using the SMFs in Figure \ref{fig:smf}, as described in Section \ref{sec:qfevol}. Darker shading indicates lower redshift and the width represents the standard jackknife uncertainty.}         \label{fig:qf}
    \end{center}
\end{figure*}

Observing the evolutionary trends in SMF for each of these
sub-populations provides a narrative of the different galaxy
evolutionary tracks involving environment and the end of star
formation. For example, the decrease in the massive star-forming
galaxies in low density environments over cosmic time can be
attributed to the transition of those galaxies to any of the other
panels. The star-forming galaxies in low density environments that
have ended star formation over time are possibly responsible
for the increase of the quiescent, low density environment SMF over
time. The star-forming galaxies that fall into higher density
environments explain the increase in the star-forming high density
environment SMF below the knee. Finally, star-forming galaxies in high
density environments that have ended their star-formation,
quiescent galaxies that have transitioned from low to high density
environments, and star-forming galaxies in low density environments
that end their star-formation while infalling to high density
environments all contribute to the overall increase of the high
environment quiescent SMF.

In addition to the evolution over cosmic time, we observe noticeable
trends when we compare the SMFs for star-forming and quiescent
galaxies between the two environments. Comparison of the SMFs in low
versus high density environments reveal a noticeable relation between
mass and density, with SMFs in high density environments having more
massive galaxies, especially evident in our lowest redshift bin. 
We further confirm this trend when we compare the
median mass between the two environments to find that the median mass
for galaxies in high density environments is significantly greater
than in low density environments. The relationship between mass and
environment observed in our SMFs reflects the well-established
mass-density relation and observed mass segregation with environment
in the literature (\citealt{norberg02a, zehavi02a, Blanton:2005ab,
  bundy06a, Scodeggio:2009aa, Bolzonella:2010aa}).

While our mass complete subsample coupled with robust environment
measurements allows us to compare SMF evolution for each of our
subsamples out to $z=0.8$, we caution readers regarding the
photometric biases affecting the SDSS imaging (and perhaps the other
imaging sources) and reserve detailed analysis of the SMFs for 
future investigation.

\section{Results: Quiescent Fraction} \label{sec:qf_const}
The SMFs calculated in the previous section illustrate the stellar
mass distribution of our galaxy population and its evolution over
cosmic time. In this section, using the SMFs of our subsamples, we
compare the quiescent and the star-forming populations by calculating
the fraction of galaxies that have ended their star-formation, the quiescent fraction. 

While the fractional relation of the star-forming and quiescent
populations has been investigated in the past, with limited
statistics, disentangling the environmental effects from underlying
correlations among observable galaxy properties such as the color-mass
or mass-density relations (\citealt{Cooper:2010aa}) remains a 
challenge. With the better statistics available from SDSS and
PRIMUS, we evaluate the quiescent fraction in bins of stellar mass,
redshift, and environment in Section \ref{sec:qfevol}. By analyzing
the quiescent fraction with respect to these properties, in Section
\ref{sec:env_qf_evol} we explicitly compare the quiescent fraction
evolution in low and high density environments. Our comparison reveal
the subtle environmental effects on the quiescent fraction
evolution. Furthermore, by quantifying this environmental effect, we
are able constrain the role of environmental effects on how galaxies
end their star formation.
\subsection{Evolution of the Quiescent Fraction} \label{sec:qfevol}
From the SMF number densities ($\Phi$) computed in the previous section, the quiescent fraction is computed as follows, 
\begin{equation}
f_{\mathrm{Q}} ( \mathcal{M}_{*}, z)= \frac{\Phi_{Q}}{\Phi_{SF}+\Phi_{Q}}.
\end{equation}
$\Phi_{Q}$ and $\Phi_{SF}$ are the total number of galaxies per unit
volume in stellar mass bin of $\Delta(\mathrm{log} \: \mathcal{M}) = 0.20
\: \mathrm{dex}$ for the quiescent and star-forming subsamples,
respectively (Equation \ref{eq:phi}). We compute $f_{\mathrm{Q}}$ for high
and low density environments for all redshift bins as plotted in
Figure \ref{fig:qf}, which shows the evolution of $f_{\mathrm{Q}}$ for
high (right panel) and low (left panel) density environments. As in
Figure \ref{fig:smf}, the evolution of the quiescent fraction over
cosmic time is represented in the shading (darker with lower redshift)
and the uncertainty is represented by the width. For the uncertainty
in the quiescent fraction, we use the standard jackknife technique,
following the same steps as for the SMF uncertainty in Section \ref{sec:smfcalc}. 

Most noticeably in Figure \ref{fig:qf}, we find $f_{\mathrm{Q}}$ increases
monotonically as a function of mass at all redshifts and
environments. In other words, for galaxies in any environment since $ z \sim 0.8$, galaxies with higher
masses are more likely to have ceased their star-formation. With the roughly linear correlation between galaxy SFR to galaxy color and morphology, we find that this trend reflects the well established color-mass and morphology-mass relations: more massive galaxies are more likely to be red or early-type (\citealt{blanton09a}). 

Focusing on the redshift evolution of $f_{\mathrm{Q}}$, we find that for
both environments $f_{\mathrm{Q}}$ increases as redshift decreases. For
high density environments, this is analogous to the Butcher-Oemler
Effect (\citealt{Butcher:1984aa}), which states that galaxy
populations in groups or clusters have higher $f_{\mathrm{blue}}$ 
(lower $f_{\mathrm{Q}}$) at higher redshift. This evolution occurs with
roughly the same amplitude in low environments as well.

In addition, when we compare the stellar masses at which $f_{\mathrm{Q}} =
0.5$ for each subsample, the so-called $\mathcal{M}_{50-50}$, we find
that this quantity decreases over cosmic time. This corresponds to the
well-known mass-downsizing pattern found by previous investigators
(e.g. \citealt{bundy06a}). Furthermore, the mass-downsizing trend
observed in each of our environment subsample is qualitatively
consistent with the trend observed in zCOSMOS Redshift Survey for
isolated and group galaxies (\citealt{Iovino:2010aa}).


Finally, we compare between our low and high density environment
$f_{\mathrm{Q}}$s at each redshift bin interval. For our lowest redshift
bin, we find that $f_{\mathrm{Q}}$ at low density environments ranges from
$\sim 0.4$ to $\sim 0.9$ for $10^{10.2} \mathcal{M}_{\odot} <
\mathcal{M}_{*} < 10^{11.5} \mathcal{M}_{\odot}$. Over the same mass
range, $f_{\mathrm{Q}}$ at high density environment ranges from $\sim
0.55$ to $\sim 0.9$. For our SDSS sample, $f_{\mathrm{Q}}$ in
high density environments is notably higher. 

For our PRIMUS sample at $z \sim 0.3$, over $10^{9.5} \mathcal{M}_{\odot} < \mathcal{M}_{*} < 10^{11} \mathcal{M}_{\odot}$ $f_{\mathrm{Q}}$ ranges from $\sim 0.2$ to $\sim 0.65$ for low density environment, while at high density environment $f_{\mathrm{Q}}$ ranges from $\sim 0.2$ to $\sim 0.8$. Similarly, at $z \sim 0.5$, over $10^{10} \mathcal{M}_{\odot} < \mathcal{M}_{*} < 10^{11.2} \mathcal{M}_{\odot}$ $f_{\mathrm{Q}}$ ranges from $\sim 0.3$ to $\sim 0.6$ for low density environment and $f_{\mathrm{Q}}$ ranges from $\sim 0.3$ to $\sim 0.7$. Finally in our highest redshift bin $z \sim 0.7$, over the mass range $10^{10.5} \mathcal{M}_{\odot} < \mathcal{M}_{*} < 10^{11.5} \mathcal{M}_{\odot}$, $f_{\mathrm{Q}}$ ranges from $\sim 0.35$ to $\sim 0.6$ for low density and $\sim 0.45$ to $\sim 0.8$ for high density. For the entire redshift range of our sample, $f_{\mathrm{Q}}$ in high density environment is higher than $f_{\mathrm{Q}}$ in low density environments. 

While there is a significant difference in $f_{\mathrm{Q}}$ between the
environments, since the difference is observed from our highest
redshift bin, it is not necessarily a result of environment dependent
mechanisms for ending star formation. In order to isolate any
environmental dependence, in the following section we quantitatively
compare the evolution of the quiescent fraction between the different
environments.

\begin{figure}
    \begin{center}
        \leavevmode
        \epsscale{1.0}
        \epsfig{file=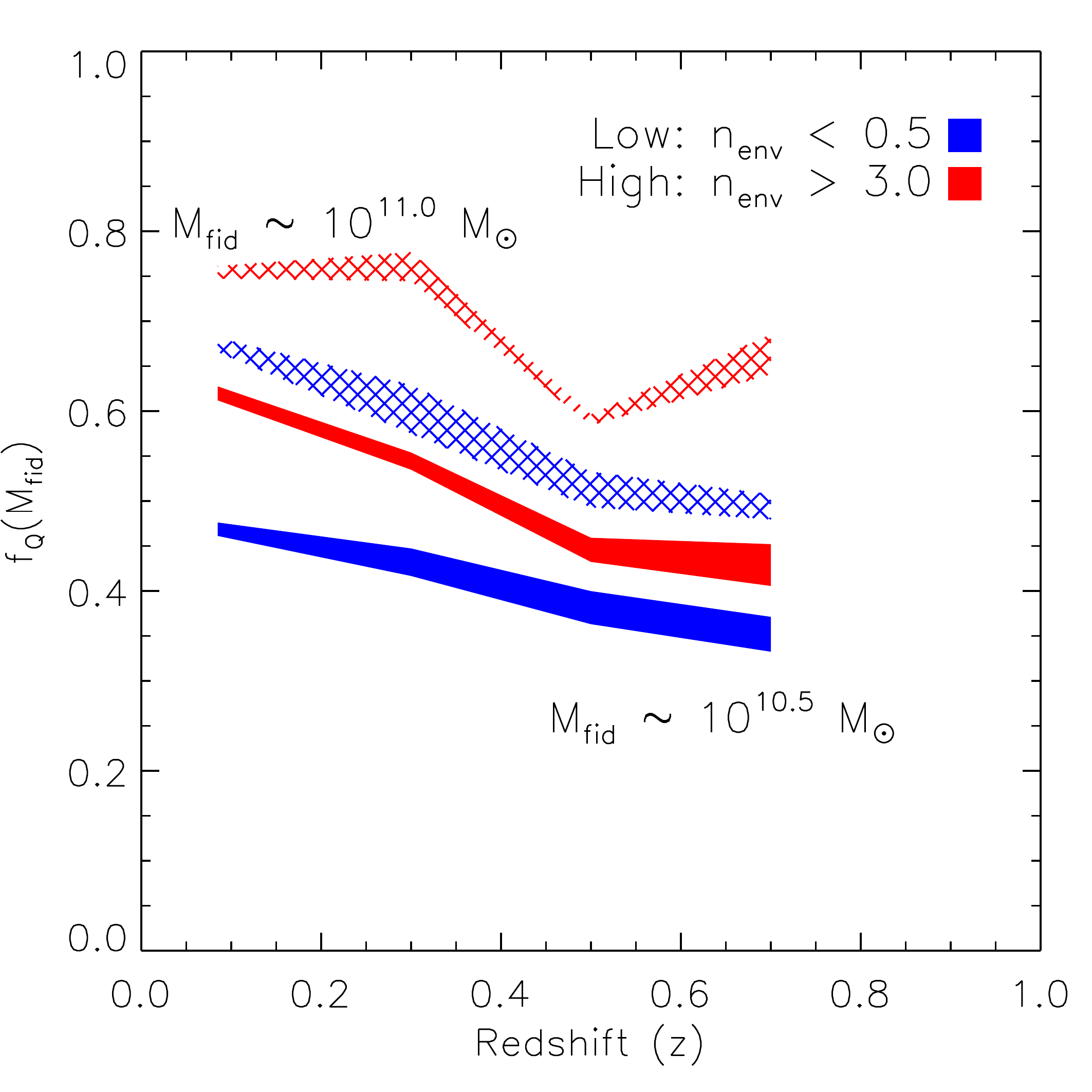, height=0.45\textwidth}
        \caption{The evolution of the quiescent fraction at fiducial
          mass, $f_{Q}(\mathcal{M}_{\mathrm{fid}})$, for low (blue) and
          high (red) density environments within the redshift range $z
          = 0.0 - 0.8$. We present the $f_{Q}(\mathcal{M}_{\mathrm{fid}})$
          evolution for $\mathcal{M}_{\mathrm{fid}} = 10^{10.5}
          \mathcal{M}_\odot$ (solid fill) and $10^{11}
          \mathcal{M}_\odot$ (patterned fill) with the uncertainty of
          the best-fit parameter $b$ in Equation \ref{eq:qffit}
          represented by the width of the line. While the high density $f_{\mathrm{Q}}(\mathcal{M}_{\mathrm{fid}})$ is greater than low density environment $f_{\mathrm{Q}}(\mathcal{M}_{\mathrm{fid}})$ over the entire redshift range of our sample, there is a significant increase in $f_{\mathrm{Q}}(\mathcal{M}_{\mathrm{fid}})$ over cosmic time for both environments. For the environment cut-offs ($n_{\mathrm{env}} < 0.5$ for low and $n_{\mathrm{env}} > 3.0$ for high), there is no significant difference in the slope of the evolution between the environments.}         \label{fig:qffit}
    \end{center}
\end{figure}

\begin{table} 
  \caption{Best Fit Parameters for $f_{\mathrm{Q}}(\mathcal{M}_{*})$ Fit}
  \label{tab:bestfitparam}
  \begin{center}
    \leavevmode
    \begin{tabular}{clcc} \hline \hline              
    $z_1 < z < z_2$ &Environment        &a  &b  \\ \hline 
$0.05 < z< 0.12$ &$n_{\mathrm{env}} < \lowenvthresh$ & $0.410 \pm 0.018$ & $0.469 \pm 0.007$ \\
               &$n_{\mathrm{env}} > \highenvthresh$ & $0.270 \pm 0.016$ & $0.620 \pm 0.008$ \\ 
                              &               &                       &                           \\ \hline   
$0.2 < z <0.4$ & $n_{\mathrm{env}} < \lowenvthresh$ & $0.340 \pm 0.032$ & $0.432 \pm 0.015$ \\
               &$n_{\mathrm{env}} > \highenvthresh$ & $0.432 \pm 0.018$ & $0.544 \pm 0.010$ \\
               &               &                       &                           \\ \hline
$0.4 < z < 0.6$      &$n_{\mathrm{env}} < \lowenvthresh$ & $0.263 \pm 0.038$ & $0.381 \pm 0.018$ \\
               &$n_{\mathrm{env}} > \highenvthresh$ & $0.289 \pm 0.018$ & $0.446 \pm 0.013$ \\
               &               &                       &                           \\ \hline
$0.6 < z < 0.8$      &$n_{\mathrm{env}} < \lowenvthresh$ & $0.284 \pm 0.036$ & $0.352 \pm 0.019$ \\
               &$n_{\mathrm{env}} > \highenvthresh$            & $0.468 \pm 0.065$ & $0.429 \pm 0.023$ \\
               &               &                       &                           \\ \hline
  \multicolumn{4}{l}{}                                             \\       
    \end{tabular} \par
    \end{center}
    {\bf Notes}: Best fit parameters in Equation \ref{eq:qffit} for each subsample $f_{\mathrm{Q}}(\mathcal{M}_{*})$ in Figure \ref{fig:qf} for $\mathcal{M}_{\mathrm{fid}} = 10^{10.5} \mathcal{M}_{\odot}$.
\end{table}

\subsection{Environmental Effects on the Quiescent Fraction Evolution} \label{sec:env_qf_evol}
In order to more quantitatively compare the $f_{\mathrm{Q}}$ evolution for different epochs and environments, we fit $f_{\mathrm{Q}}$ for each subsample to a power-law parameterization as a function of stellar mass, 
\begin{equation} \label{eq:qffit}
f_{\mathrm{Q}}(\mathcal{M}_{*}) = a \: \mathrm{log} \; \left(\frac{ \mathcal{M}_{*}}{\mathcal{M}_{\mathrm{fid}}} \right)+b,
\end{equation}
where $a$ and $b$ are best-fit parameters using {\em MPFIT} (\citealt{Markwardt:2009aa}) and $\mathcal{M}_{\mathrm{fid}}$ represents the empirically selected fiducial mass within the stellar mass limits where there is a sufficiently large number of galaxies. We primarily focus on $\mathcal{M}_{\mathrm{fid}} = 10^{10.5} \: \mathcal{M}_{\odot}$. 

In Figure \ref{fig:qffit} we present the evolution of
$f_{\mathrm{Q}}(\mathcal{M}_{\mathrm{fid}})$ from $z \sim 0.7$ to $\sim 0.1$
at low (blue) and high (red) density environments for
$\mathcal{M}_{\mathrm{fid}} = 10^{10.5} \: \mathcal{M}_{\odot}$ (solid
fill) and $10^{11} \: \mathcal{M}_{\odot}$ (pattern fill). The width
of the evolution represents the uncertainty derived from {\em
  MPFIT}. As noted earlier in Section \ref{sec:qfevol}, $f_{\mathrm{Q}}$
in high density environments is significantly greater than
$f_{\mathrm{Q}}$ in low density environments for both fiducial mass
choices. Throughout our sample's redshift range
$f_{\mathrm{Q}}(\mathcal{M}_{\mathrm{fid}})_{\mathrm{high}} -
f_{\mathrm{Q}}(\mathcal{M}_{\mathrm{fid}})_{\mathrm{low}} \sim 0.1$.

In addition, the $f_{\mathrm{Q}}(\mathcal{M}_{\mathrm{fid}})$
evolution illustrates that the quiescent fraction in low density
environment increases over cosmic time:
$f_{\mathrm{Q}}(\mathcal{M}_{\mathrm{fid}}, z \sim 0.1) -
f_{\mathrm{Q}}(\mathcal{M}_{\mathrm{fid}}, z \sim 0.7) \sim 0.1$. This
significant quiescent fraction evolution for low density environments
suggests that internal mechanisms, independent of environment, are
responsible for a significant amount of star-formation cessation. Meanwhile, the $f_{\mathrm{Q}}(\mathcal{M}_{\mathrm{fid}})$ evolution in high density environment ($f_{\mathrm{Q}}(\mathcal{M}_{\mathrm{fid}}, z \sim 0.1) - f_{\mathrm{Q}}(\mathcal{M}_{\mathrm{fid}}, z \sim 0.7) \sim 0.12$) shows little additional evolution.

When we increase our choice of $\mathcal{M}_{\mathrm{fid}}$ to $10^{11}
\mathcal{M}_\odot$, aside from an overall shift in
$f_{\mathrm{Q}}(\mathcal{M}_{\mathrm{fid}})$ by $\sim 0.2$, we observe the
same evolutionary trends. $f_{\mathrm{Q}}(\mathcal{M}_{\mathrm{fid}} =
10^{11}\mathcal{M}_{\odot})$ for both low and high density
environments each increase by $\sim 0.2$ from at all redshifts we
study. Increasing the fiducial mass to $10^{11}
\mathcal{M}_{\odot}$ does not significantly alter the evolutionary
trends in either environment. Although the varying stellar mass
completeness at each redshift bin limits the masses we probe for the
$f_{\mathrm{Q}}$ evolution, our $f_{\mathrm{Q}}$ evolution exhibits little
mass dependence.

Because our fixed aperture definition of environment is susceptible to
contamination due to PRIMUS redshift errors, we consider in Figure
\ref{fig:qffit_comp} more stringent high density environment
classifications, extending the cut off to $n_{\mathrm{env}} > 5$ and
$7$ (specified in the top right legend and represented by the color of
the shading). Aside from the increase in uncertainties that accompany
the decrease in sample size of the purer high environment sample, we
find an extension of the $f_{\mathrm{Q}}$ difference between the
environments we stated earlier. A more stringent high environment
classification significantly increases the overall
$f_{\mathrm{Q}}(\mathcal{M}_{\mathrm{fid}})$, which rises monotonically with
the $n_{\mathrm{env}}$ limit.

More importantly, a purer high environment classification reveals a
more significant environment dependence on the $f_{\mathrm{Q}}$
evolution. While the difference between the $f_{\mathrm{Q}}$ evolution in
low and high density environment is negligible for the $n_{\mathrm{env}} >
3$ cut-off, there is a notable difference in $f_{\mathrm{Q}}$
evolution between our highest cut-off $n_{\mathrm{env}} > 7$ and our low
density environment. $f_{\mathrm{Q}}(\mathcal{M}_{\mathrm{fid}}, z
\sim 0.1) - f_{\mathrm{Q}}(\mathcal{M}_{\mathrm{fid}}, z \sim 0.7)
\sim 0.25$ for $n_{\mathrm{env}} > 7$ versus
$f_{\mathrm{Q}}(\mathcal{M}_{\mathrm{fid}}, z \sim 0.1) -
f_{\mathrm{Q}}(\mathcal{M}_{\mathrm{fid}}, z \sim 0.7) \sim 0.1$ for
low density environment. In addition to the
environment independent internal mechanisms that can explain the
$f_{\mathrm{Q}}$ evolution in low density environments, there may be other
environment dependent mechanisms that can account for the moderate
environment dependence of the $f_{\mathrm{Q}}$ evolution. Our measured
difference in the $f_{\mathrm{Q}}$ evolution between environments provides
an important constraint for any environmental models for ending star
formation.

\def \iovinopanel {b}
\def \kovacpanel {b}
\def \pengpanel {c}
\begin{figure*}
    \begin{center}
        \leavevmode
        \epsscale{1.0}
        \epsfig{file=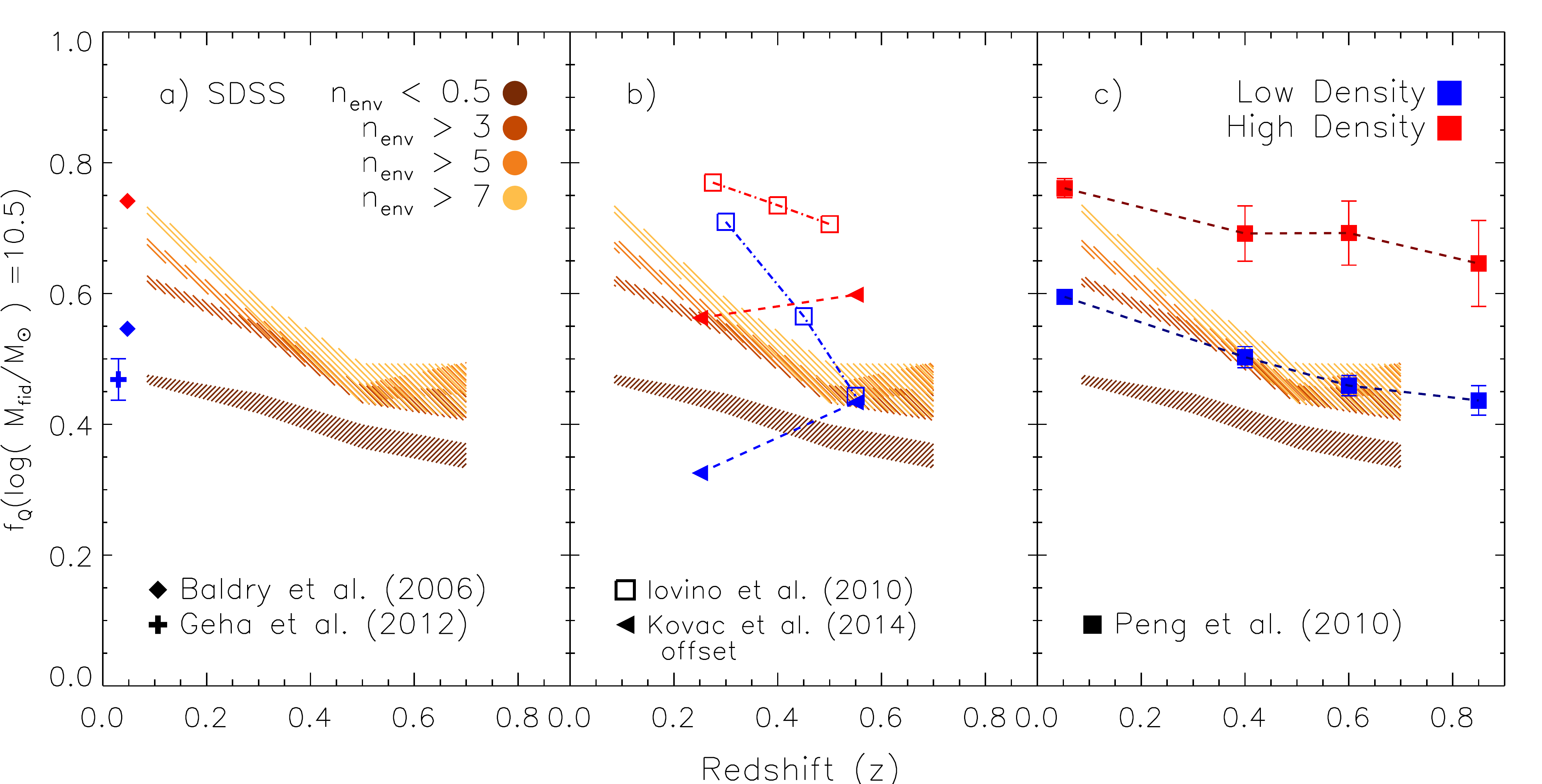, height=\textwidth, angle=90}
        \caption{$f_{\mathrm{Q}}(\mathcal{M}_{\mathrm{fid}}=10^{10.5} \mathcal{M}_\odot)$ evolution compared to $f_{\mathrm{red}}(\mathcal{M}_{*} \sim 10^{10.5} \; \mathcal{M}_{\odot})$ in the literature: \cite{Baldry:2006aa} (diamond) and \cite{geha12a} (cross) from SDSS (panel a), \cite{Iovino:2010aa} (empty square) and \cite{Kovac:2014aa} (triangle) from zCOSMOS (panel \kovacpanel), and \cite{Peng:2010aa} from both SDSS and zCOSMOS (panel \pengpanel). The $f_{\mathrm{red}}$ values from \cite{Iovino:2010aa}, \cite{Kovac:2014aa}, \cite{Baldry:2006aa}, and \cite{Peng:2010aa} are calculated from the best-fit parameterizations presented in the respective works. High density environment is represented in red and low density environment is represented in blue. The $f_{\mathrm{Q}}$ value from \cite{geha12a} is the $f_{\mathrm{Q}}$ value at $\mathcal{M} = 10^{10.55} \mathcal{M}_{\odot}$. Uncertainties in the \cite{Iovino:2010aa} best-fit $f_{\mathrm{red}}$ is omitted due to insufficient information on the cross correlation terms of the fit parameters. For \cite{Kovac:2014aa} we apply the offset between the color-based and SFR-based galaxy classification in order to plot the $f_{\mathrm{Q}}$ estimates. We also plot the $f_{\mathrm{Q}}(\mathcal{M}_{\mathrm{fid}} = 10^{10.5} \mathcal{M}_\odot)$ evolution of our sample with varying environment cut-offs specified on the top right. As in Figure \ref{fig:qffit} the width of the $f_{\mathrm{Q}}(\mathcal{M}_{\mathrm{fid}} = 10^{10.5} \mathcal{M}_\odot)$ evolution represent the uncertainty in the best-fit parameters of Equation \ref{eq:qffit}.}         \label{fig:qffit_comp}
    \end{center}
\end{figure*}

\subsection{Comparison to Literature}
Although a direct comparison with other results is difficult due to
our sample specific methodology, a number of results from the literature have investigated the quiescent fraction in comparable fashions. In this section we compare our $f_{\mathrm{Q}}$ results from above to a number of these results, specifically from SDSS and zCOSMOS, with similarly defined samples and analogous environment classifications. 

In Figure \ref{fig:qffit_comp}, we plot  best-fit parameterization of
$f_{\mathrm{red}}$ for high and low density environment from SDSS (panel a), zCOSMOS (panel \iovinopanel), and \cite{Peng:2010aa} (filled square; panel d) from both surveys. 
From \cite{Iovino:2010aa} (empty square; panel \iovinopanel), we calculate $f_{\mathrm{red}} = 1-f_{\mathrm{blue}}$
using the best-fit $f_{\mathrm{blue}}$ from the mass bin $\mathcal{M} =
10^{10.3} - 10^{10.8} \mathcal{M}_{\odot}$. From \cite{Kovac:2014aa} (triangle; panel \kovacpanel)
we plot an estimated $f_{\mathrm{Q}}$ by applying the residual between SFR
based and color based galaxy classifications to the best-fit
$f_{\mathrm{red}}$ at $\mathcal{M} = 10^{10.5} \mathcal{M}_{\odot}$ for
low ($\delta = 0.0$) and high density environments ($\delta =
1.5$). Similarly, from \cite{Baldry:2006aa} (diamond; panel a) we plot $f_{\mathrm{Q}}$
derived from the best-fit $f_{\mathrm{red}}$ at $\mathcal{M} = 10^{10.5}
\mathcal{M}_{\odot}$ for low ($\delta = 0.0$) and high density
environment ($\delta = 1.0$). For \cite{geha12a} (cross; panel a), we plot $f_{\mathrm{Q}}$
for their isolated galaxy sample in their mass bin closest to $10^{10.5} \mathcal{M}_{\odot}$, $\mathcal{M} = 10^{10.55} \mathcal{M}_{\odot}$. Finally for \cite{Peng:2010aa} (square; panel \pengpanel), we plot the parameterized $f_{\mathrm{red}}$ at $\mathcal{M} = 10^{10.5} \mathcal{M}_{\odot}$ using their best-fit parameters for low ($\delta =0.0$) and high ($\delta = 1.4$) density environments. 

For our lowest redshift bin SDSS sample, we find that our $f_{\mathrm{Q}}$
for low and high environments are consistent with other SDSS
$f_{\mathrm{Q}}$ (or $f_{\mathrm{red}}$) measurements as a function of
environment. For example, \cite{Baldry:2006aa} uses projected neighbor
density environment measures ($\mathrm{log} \; \Sigma$) to obtain
$f_{\mathrm{Q}}(\mathcal{M})$ for a range of environmental
densities. Although the different environment measurements make direct
comparisons difficult, in their corresponding higher environments
($\mathrm{log} \; \Sigma > 0.2$ in \citealt{Baldry:2006aa})
$f_{\mathrm{Q}}(\mathcal{M} \sim 10^{10.2} \mathcal{M}_{\odot}) \sim 0.6$
and $f_{\mathrm{Q}}(\mathcal{M} \sim 10^{11.5} \mathcal{M}_{\odot}) \sim
0.9$, which is in agreement with our high density
environment. Likewise, for lower environments ($\mathrm{log} \; \Sigma <
-0.4$ in \citealt{Baldry:2006aa}) $f_{\mathrm{Q}}(\mathcal{M} \sim
10^{10.2} \mathcal{M}_{\odot}) \sim 0.4$ and $f_{\mathrm{Q}}(\mathcal{M}
\sim 10^{11.5} \mathcal{M}_{\odot}) \sim 0.8$, which also agree with
our low density environment $f_{\mathrm{Q}}$. The \cite{Baldry:2006aa}
points (diamond) in Figure \ref{fig:qffit_comp} reflect this
agreement.

More recently, \cite{Tinker:2011aa}, using a group-finding algorithm on the SDSS DR7, presents the relationship between $f_{\mathrm{Q}}$ and overdensity for galaxies within the mass range $\mathrm{log} \; \mathcal{M} = [9.8, 10.1]$. The \cite{Tinker:2011aa} $f_{\mathrm{Q}}$ at the lowest and highest overdensities, $f_{\mathrm{Q}} \sim 0.4$ and $f_{\mathrm{Q}} \sim 0.6$ respectively, are consistent with our $f_{\mathrm{Q}}$ for low and high density environment at the lower mass limit ($\mathrm{log} \; \mathcal{M} \sim 10.2$). 

A modified \cite{Tinker:2011aa} sample is used in \cite{geha12a} to
obtain $f_{\mathrm{Q}}$ for isolated galaxies over a wider mass range
($10^{7.4} \mathcal{M}_\odot$ to $10^{11.2}
\mathcal{M}_{\odot}$). Although \cite{geha12a} probe a slightly lower
redshift range ($ z \le 0.06$), their $f_{\mathrm{Q}}$ is consistent with
our low density sample. Within the overlapping mass range, at the low
mass end \cite{geha12a} find $f_{\mathrm{Q}}(\mathcal{M}_{*} \sim
10^{10.2} \mathcal{M}_{\odot}) \sim 0.3$ and at the high mass end they
find $f_{\mathrm{Q}}(\mathcal{M}_{*} \sim 10^{11.2} \mathcal{M}_{\odot})
\sim 0.8$. Both of these values agree with our lowest redshift
$f_{\mathrm{Q}}$ results in low density environment. Figure
\ref{fig:qffit_comp} illustrates the $f_{\mathrm{Q}}$ agreement for
$\mathcal{M}_{*} = 10^{10.5} \mathcal{M}_{\odot}$.


For $z > 0.2$, we compare our PRIMUS $f_{\mathrm{Q}}$ results to the $f_{\mathrm{red}}$ (or $1-f_{\mathrm{blue}}$) results from the zCOSMOS Redshift Survey (\citealt{Iovino:2010aa, Kovac:2014aa}), which covers a similar redshift range as PRIMUS. \cite{Iovino:2010aa}, and \cite{Kovac:2014aa} using a mass-complete galaxy sample derived from zCOSMOS and a group catalog, 3D local density contrast, and overdensity environment measurements, respectively, compare $f_{\mathrm{red}}$ with respect to environment. The $f_{\mathrm{blue}}$ for group and isolated galaxies from \cite{Iovino:2010aa} are generally inconsistent with our $1-f_{\mathrm{Q}}$ for high and low density environments. 

Similarly, $f_{\mathrm{red}}$ for high and low overdensities in
\cite{Kovac:2014aa} are  greater overall than the PRIMUS $f_{\mathrm{Q}}$
values in high and low density environments. However,
\cite{Kovac:2014aa} points out that there is a significant difference
between classifying the quiescent population using color and SFR due to dust-reddening in star-forming galaxies. For their lower redshift bin ($0.1 < z < 0.4$) \cite{Kovac:2014aa} find that their $f_{\mathrm{Q}}$ defined by color is greater than $f_{\mathrm{Q}}$ defined by SFR by roughly $0.2$. While for their higher redshift bin ($0.4 < z < 0.7$) the difference is $0.15-0.19$. Although \cite{Kovac:2014aa} does not elaborate on how the galaxy classification discrepancy applies to the different environments, if we simply account for the difference uniformly for $f_{\mathrm{red}}$ at all environments, the \cite{Kovac:2014aa} results in their lower redshift bin are roughly consistent with our $f_{\mathrm{Q}}$ at high and low density environments. Even accounting for the dust-reddening of $f_{\mathrm{red}}$, \cite{Kovac:2014aa} finds a significantly higher $f_{\mathrm{Q}}$ in their higher redshift bin. 

In Figure \ref{fig:qf}, the $f_{\mathrm{Q}}$ evolution with respect to
mass reveals, qualitatively, little mass dependence in the
evolution. Moreover, in Figure \ref{fig:qffit}, we illustrated that
adjusting the fiducial mass only shifted the overall
$f_{\mathrm{Q}}(\mathcal{M}_{\mathrm{fid}})$, but did not change the
$f_{\mathrm{Q}}$ evolutionary trend. The consistency in the $f_{\mathrm{Q}}$
evolutionary trends over change in fiducial mass suggests that
$f_{\mathrm{Q}}$ evolution exhibit little mass dependence within the 
mass range probed in our analysis. In contrast to
the weak mass dependence we observe in our results,
\cite{Iovino:2010aa} find significantly different $f_{\mathrm{Q}}$
evolution at $\mathcal{M} \sim 10^{11} \mathcal{M}_{\odot}$ and
$\mathcal{M} \sim 10^{10.5} \mathcal{M}_{\odot}$, for both group and
isolated galaxies. In fact at their highest mass bin ($10^{10.9} -
10^{11.4} \mathcal{M}_{\odot}$), \cite{Iovino:2010aa} find no evolution
for both environments: constant $f_{\mathrm{blue}} \sim 0.1$ over $z =
0.3 - 0.8$ for both group and isolated galaxy populations.

Meanwhile in their mass bin most comparable to $\mathcal{M}_{\mathrm{fid}}
\sim 10^{10.5} \mathcal{M}_{\odot}$ ($10^{10.3} \mathcal{M}_{\odot} -
10^{10.8} \mathcal{M}_{\odot}$), \cite{Iovino:2010aa} finds that
$f_{\mathrm{blue}}$ evolves by $\sim 0.1$ from $z = 0.5$ to $0.25$ for
group galaxies and by $\sim 0.3$ from $z=0.55$ to $0.3$ for isolated
galaxies as presented in panel (\iovinopanel) of Figure \ref{fig:qffit_comp}. Altogether, with
mass bins beyond the fiducial masses we explore, \cite{Iovino:2010aa}
find a strong mass dependence with $f_{\mathrm{Q}}$ evolving significantly
more in lower mass bins. While our sample from PRIMUS provides larger
statistics than zCOSMOS, the mass-completeness limits we impose on our
sample limits the mass range we probe (e.g. $\mathcal{M} > 10^{10.5}
\mathcal{M}_{\odot}$ for our $z \sim 0.7$ bin). Consequently our
results cannot rule out mass dependence in the $f_{\mathrm{Q}}$ evolution
at lower masses.

In Figure \ref{fig:qffit} and Figure \ref{fig:qffit_comp} we
quantified that throughout our redshift range, high density
environments have a significantly greater
$f_{\mathrm{Q}}(\mathcal{M}_{\mathrm{fid}})$ than the low density
environments. This finding is in agreement with the zCOSMOS results
from \cite{Cucciati:2010aa} and \cite{Kovac:2014aa}. As illustrated in
panel (\kovacpanel) of Figure \ref{fig:qffit_comp}, \cite{Kovac:2014aa} finds $f_{\mathrm{Q}}$ in high density environment
significantly greater than $f_{\mathrm{Q}}$ at low density
environment. Moreover, since galaxy color serves as a proxy for SFR,
our results support the existence of the color-density relation
(\citealt{Cucciati:2010aa, Cooper:2010aa}) and is not consistent with
the color-density relation being merely a reflection of the
mass-density relationship, as \cite{Scodeggio:2009aa} suggest it is
based on the Vimos VLT Deep Survey ($0.2 < z< 1.4$).


In Section \ref{sec:env_qf_evol}, we showed that $f_{\mathrm{Q}}$ in
low density environments evolves over cosmic time. From this trend we
deduce that internal, environment independent, mechanisms contribute
to ending star-formation in galaxy evolution. \cite{Iovino:2010aa}
from zCOSMOS, plotted in Figure \ref{fig:qffit_comp}  panel (\iovinopanel), also find that
$f_{\mathrm{Q}}$ in low density environment increases with decreasing redshift. On the other hand \cite{Kovac:2014aa}, also from zCOSMOS, presents that $f_{\mathrm{Q}}$ in low density environment decreases over cosmic time. While the uncertainties for the parameterized $f_{\mathrm{Q}}$ are not listed, and thus not shown in Figure \ref{fig:qffit_comp}, once they are accounted for, \cite{Kovac:2014aa} find no significant $f_{\mathrm{Q}}$ evolution over cosmic time. However, once we account for the dust-reddening of the $f_{\mathrm{red}}$, we find a more significant decrease over cosmic time (Figure \ref{fig:qffit_comp} panel \kovacpanel).

Furthermore, in Section \ref{sec:env_qf_evol}, our comparison of the
$f_{\mathrm{Q}}$ evolution between the lowest density environment and the
highest density environment revealed a modicum of evidence for the
existence of environment dependent mechanisms. The same comparison
with zCOSMOS results (\citealt{Iovino:2010aa, Kovac:2014aa}) present
trends inconsistent with our findings. First, comparing the high (red)
and low (blue) density environments for \cite{Iovino:2010aa} in Figure
\ref{fig:qffit_comp} shows that there are indeed pronounced
discrepancies between the $f_{\mathrm{Q}}$ evolution in different
environments. Group galaxies in \cite{Iovino:2010aa} have higher
overall $f_{\mathrm{Q}}$ than isolated galaxies. However, unlike our
results, which find a greater $f_{\mathrm{Q}}$ evolution at higher density
environments, \cite{Iovino:2010aa} finds the opposite environment
dependence that there is a significantly greater $f_{\mathrm{Q}}$
evolution for isolated galaxies.

Next, \cite{Kovac:2014aa} also find that overall $f_{\mathrm{Q}}$ is
greater in high density than in low density environments. Like their
low density environment $f_{\mathrm{Q}}$ evolution, $f_{\mathrm{Q}}$ in high
density environment decreases over cosmic time between their two
redshift bins. Although the decrease in $f_{\mathrm{Q}}$ over cosmic time
conflicts with our results, \cite{Kovac:2014aa} finds a greater (less
negative) $f_{\mathrm{Q}}$ evolution in high density environments relative
to low density environments, suggesting an environment dependence that
is in the same direction as our results. We note that the negative
slopes of the $f_{\mathrm{Q}}$ evolution in both environments are enhanced
in Figure \ref{fig:qffit_comp} due to the dust-reddening
correction we impose to the \cite{Kovac:2014aa} $f_{\mathrm{red}}$ results. 


While the zCOSMOS survey provides more precise spectroscopic redshifts than PRIMUS, our sample provides significantly larger statistics. In addition, our sample covers a larger portion of the sky. Our SDSS-{\em GALEX} sample covers $2,505 \;\mathrm{deg}^2$. More comparably, our PRIMUS sample covers $5.5 \;\mathrm{deg}^2$, over 3 times the sky coverage of zCOSMOS ($1.7 \;\mathrm{deg}^2$). Furthermore, our PRIMUS sample is constructed from five independent fields which allows us to significantly reduce the effects of cosmic variance. 

As listed in Table \ref{tab:subsample}, after our edge effect cuts and
stellar mass completeness limits, our sample consists of $13,734$
galaxies from PRIMUS over $0.2< z< 0.8$ and $63,417$ galaxies from
SDSS over $0.05 < z < 0.12$. Meanwhile, \cite{Iovino:2010aa} has $914$
galaxies with $\mathcal{M} > 10^{10.3} \mathcal{M}_{\odot}$ over $0.1
< z < 0.6$ and $1033$ galaxies with $\mathcal{M} > 10^{10.6}
\mathcal{M}_{\odot}$ over $0.1 < z < 0.8$. For the actual sample used
to obtain the best-fit $f_{\mathrm{Q}}$ values in Figure
\ref{fig:qffit_comp} \cite{Iovino:2010aa} has $617$ galaxies. In
comparison, our PRIMUS sample alone contains $> 20$ times the number
of galaxies. While there is a considerable difference in the overall
$f_{\mathrm{Q}}$ between our results and those of \cite{Iovino:2010aa},
the use of different methodologies, particularly for galaxy
classification and environment measurements, make such comparisons
ambiguous. On the other hand, the discrepancies in the $f_{\mathrm{Q}}$
evolutionary trends with our results may be explained by the limited
statistics in the \cite{Iovino:2010aa} sample.

The more recent \cite{Kovac:2014aa} provides larger statistics with
$2,340$ galaxies in their lower redshift bin ($0.1 < z < 0.4$) and
$2,448$ galaxies in their higher redshift bin ($0.4 < z <
0.7$). Although their sample is smaller than the PRIMUS sample, which
contains over twice times the number of galaxies, the
\cite{Kovac:2014aa} sample provides a more stable comparison.  Once
their results are adjusted for the dust-reddening, we
find that their overall $f_{\mathrm{Q}}$ is more or less consistent with
our overall $f_{\mathrm{Q}}$. However, it is difficult to explain the
significant discrepancies in the $f_{\mathrm{Q}}$ evolutionary trends.
The significant overdenities observed in the COSMOS field at $z \sim 0.35$ and $z \sim 0.7$ (\citealt{Lilly:2009aa, Kovac:2010ab}) may have a significant effect on the zCOSMOS results and offer a possible explanation for the discrepancies. 

\section{Summary and Discussion} \label{sec:summary}
Using a stellar mass complete galaxy sample derived from SDSS and
PRIMUS accompanied by a consistently measured galaxy environment from
robust spectroscopic redshifts, we measure the stellar mass functions
for star-forming and quiescent galaxies in low and high density
environments over the redshift range $0.05 < z < 0.8$. From these
stellar mass functions, we compare the proportion of galaxies that
have ended their star-formation within the subsamples by computing the
quiescent fraction for each of them. In order to better quantify the
evolution of the quiescent fraction over cosmic time, we fit our
quiescent fraction anchored at a fiducial mass. 

From our analyses we find the following notable results. The first
three demonstrate that previous findings that are well known in the
local universe are applicable out to $z\sim 0.7$. The last two are
consistent with the findings of \cite{Peng:2010aa} but provide increased
detail on the environmental dependence of galaxy evolution:
\begin{enumerate}
	\item From the SMFs, we find that the galaxy population in high
    density environments, both star-forming and quiescent, have a
    higher median mass, thus confirming the mass-density relation and
    mass-segregation in different environments throughout our sample's
    redshift range.
	\item For all subsamples, $f_{\mathrm{Q}}$ increases monotonically with
    galaxy stellar mass, showing a clear mass dependence and
    reflecting the well-established color-mass and morphology-mass
    relations.
	\item We illustrate that $f_{\mathrm{Q}}$ in high density environments
    is greater than $f_{\mathrm{Q}}$ in low density environments
    regardless of mass and out to redshift $z\sim 0.7$. This result
    reflects the well known trend that galaxies in high density
    environment are statistically redder, have lower SFRs, and are
    more massive.
	\item $f_{\mathrm{Q}}$ increases significantly with redshift for both
    low and high density environments. For high density environment,
    this trend is the Butcher-Oemler effect. Furthermore, the
    $f_{\mathrm{Q}}$ evolution in low density environment suggest the
    existence of internal environment-independent mechanisms for
    ending star formation.
	\item Comparison of the $f_{\mathrm{Q}}(\mathcal{M}_{\mathrm{fid}})$
    evolution for a range of environment classifications reveals that
    the since $z = 0.8$, $f_{\mathrm{Q}}$ has evolved by a greater amount
    in the highest density environments. For our purest high
    environment sample ($n_{\mathrm{env}} > 7$), the total $f_{\mathrm{Q}}$
    evolution is $\sim 0.1$ greater than the total $f_{\mathrm{Q}}$
    evolution in low density environment, revealing a moderate
    dependence on environment.
\end{enumerate}

Many physical mechanisms have been proposed to explain the cessation
of star-formation observed in many galaxies. Recently star-formation
cessation has often been classified into internal or external
mechanisms, and sometimes more specifically into mass-dependent and
environment-dependent mechanisms (\citealt{Baldry:2006aa,
  Peng:2010aa}). The significant redshift evolution of the
$f_{\mathrm{Q}}$ in low density environments confirms the existence of
internal mechanisms that end star-forming in galaxies.

Furthermore, the greater $f_{\mathrm{Q}}$ evolution in the highest density
environment relative to low density environments suggests that in
addition to the internal mechanisms, in high density environments such
as groups and clusters, environment-dependent effects may also
contribute to the end of star-formation. Our results do not
specifically shed light on which mechanisms (e.g. strangulation,
ram-pressure stripping, etc.) occur in high density environments. Not to 
mention, the mechanism could yet be indirect; for example, the
galaxies in higher density environments could end star-formation
primarily due internal processes, which affect the galaxies that end up
in groups and clusters more greatly.  Nevertheless, our results impose
important constraints on the total possible contribution of
environment dependent mechanisms that models must satisfy, providing a
limit on the role of environment in ending star formation in
galaxies. 

\bigskip 
CH and MB were supported by NSF-AST-1109432 and NSF-AST-0908354. MB was also supported by NSF-AST-1211644. ALC acknowledges support from NSF CAREER award. We acknowledge Katarina Kova\u{c} and Ying-Jie Peng for  insightful discussions. We also thank Marla Geha for providing published quiescent fraction results in electronic format.

Funding for PRIMUS has been provided by NSF grants AST-0607701, 0908246, 0908442, 0908354, and NASA grant 08-ADP08-0019. The Galaxy Evolution Explorer (GALEX) is a NASA Small Explorer, launched in April 2003, whose mission was developed in cooperation with the Centre National d`Etudes Spatiales of France and the Korean Ministry of Science and Technology. Funding for the SDSS and SDSS-II has been provided by the Alfred P. Sloan Foundation, the Participating Institutions, the National Science Foundation, the U.S. Department of Energy, the National Aeronautics and Space Administration, the Japanese Monbukagakusho, the Max Planck Society, and the Higher Education Funding Council for England. The SDSS Web Site is http://www.sdss.org. The SDSS is managed by the Astrophysical Research Consortium for the Participating Institutions. 

%
%
\bibliographystyle{apj}

\begin{thebibliography}{}
\expandafter\ifx\csname natexlab\endcsname\relax\def\natexlab#1{#1}\fi

\bibitem[{{Abazajian} {et~al.}(2009){Abazajian}, {Adelman-McCarthy},
  {Ag{\"u}eros}, {Allam}, {Allende Prieto}, {An}, {Anderson}, {Anderson},
  {Annis}, {Bahcall}, \& et~al.}]{Abazajian:2009aa}
{Abazajian}, K.~N., {Adelman-McCarthy}, J.~K., {Ag{\"u}eros}, M.~A., {et~al.}
  2009, \apjs, 182, 543

\bibitem[{{Baldry} {et~al.}(2006){Baldry}, {Balogh}, {Bower}, {Glazebrook},
  {Nichol}, {Bamford}, \& {Budavari}}]{Baldry:2006aa}
{Baldry}, I.~K., {Balogh}, M.~L., {Bower}, R.~G., {et~al.} 2006, \mnras, 373,
  469

\bibitem[{{Baldry} {et~al.}(2008){Baldry}, {Glazebrook}, \&
  {Driver}}]{Baldry:2008aa}
{Baldry}, I.~K., {Glazebrook}, K., \& {Driver}, S.~P. 2008, \mnras, 388, 945

\bibitem[{{Balogh} {et~al.}(2000){Balogh}, {Navarro}, \&
  {Morris}}]{Balogh:2000aa}
{Balogh}, M.~L., {Navarro}, J.~F., \& {Morris}, S.~L. 2000, \apj, 540, 113

\bibitem[{{Behroozi} {et~al.}(2013){Behroozi}, {Marchesini}, {Wechsler},
  {Muzzin}, {Papovich}, \& {Stefanon}}]{Behroozi:2013aa}
{Behroozi}, P.~S., {Marchesini}, D., {Wechsler}, R.~H., {et~al.} 2013, \apjl,
  777, L10

\bibitem[{{Bekki}(2009)}]{Bekki:2009aa}
{Bekki}, K. 2009, \mnras, 399, 2221

\bibitem[{{Bell} {et~al.}(2004){Bell}, {Wolf}, {Meisenheimer}, {Rix}, {Borch},
  {Dye}, {Kleinheinrich}, {Wisotzki}, \& {McIntosh}}]{Bell:2004aa}
{Bell}, E.~F., {Wolf}, C., {Meisenheimer}, K., {et~al.} 2004, \apj, 608, 752

\bibitem[{{Bell} {et~al.}(2005){Bell}, {Papovich}, {Wolf}, {Le Floc'h},
  {Caldwell}, {Barden}, {Egami}, {McIntosh}, {Meisenheimer},
  {P{\'e}rez-Gonz{\'a}lez}, {Rieke}, {Rieke}, {Rigby}, \& {Rix}}]{bell05a}
{Bell}, E.~F., {Papovich}, C., {Wolf}, C., {et~al.} 2005, \apj, 625, 23

\bibitem[{{Bernardi} {et~al.}(2007){Bernardi}, {Hyde}, {Sheth}, {Miller}, \&
  {Nichol}}]{Bernardi:2007aa}
{Bernardi}, M., {Hyde}, J.~B., {Sheth}, R.~K., {Miller}, C.~J., \& {Nichol},
  R.~C. 2007, \aj, 133, 1741

\bibitem[{{Blanton} {et~al.}(2006){Blanton}, {Eisenstein}, {Hogg}, \&
  {Zehavi}}]{Blanton:2006aa}
{Blanton}, M.~R., {Eisenstein}, D., {Hogg}, D.~W., \& {Zehavi}, I. 2006, \apj,
  645, 977

\bibitem[{{Blanton} {et~al.}(2011){Blanton}, {Kazin}, {Muna}, {Weaver}, \&
  {Price-Whelan}}]{Blanton:2011aa}
{Blanton}, M.~R., {Kazin}, E., {Muna}, D., {Weaver}, B.~A., \& {Price-Whelan},
  A. 2011, \aj, 142, 31

\bibitem[{{Blanton} {et~al.}(2005{\natexlab{a}}){Blanton}, {Lupton},
  {Schlegel}, {Strauss}, {Brinkmann}, {Fukugita}, \&
  {Loveday}}]{Blanton:2005ab}
{Blanton}, M.~R., {Lupton}, R.~H., {Schlegel}, D.~J., {et~al.}
  2005{\natexlab{a}}, \apj, 631, 208

\bibitem[{{Blanton} \& {Moustakas}(2009)}]{blanton09a}
{Blanton}, M.~R., \& {Moustakas}, J. 2009, \araa, 47, 159

\bibitem[{{Blanton} \& {Roweis}(2007)}]{Blanton:2007aa}
{Blanton}, M.~R., \& {Roweis}, S. 2007, \aj, 133, 734

\bibitem[{{Blanton} {et~al.}(2005{\natexlab{b}}){Blanton}, {Schlegel},
  {Strauss}, {Brinkmann}, {Finkbeiner}, {Fukugita}, {Gunn}, {Hogg},
  {Ivezi{\'c}}, {Knapp}, {Lupton}, {Munn}, {Schneider}, {Tegmark}, \&
  {Zehavi}}]{Blanton:2005aa}
{Blanton}, M.~R., {Schlegel}, D.~J., {Strauss}, M.~A., {et~al.}
  2005{\natexlab{b}}, \aj, 129, 2562

\bibitem[{{Bolzonella} {et~al.}(2010){Bolzonella}, {Kova{\v c}}, {Pozzetti},
  {Zucca}, {Cucciati}, {Lilly}, {Peng}, {Iovino}, {Zamorani}, {Vergani},
  {Tasca}, {Lamareille}, {Oesch}, {Caputi}, {Kampczyk}, {Bardelli}, {Maier},
  {Abbas}, {Knobel}, {Scodeggio}, {Carollo}, {Contini}, {Kneib}, {Le
  F{\`e}vre}, {Mainieri}, {Renzini}, {Bongiorno}, {Coppa}, {de la Torre}, {de
  Ravel}, {Franzetti}, {Garilli}, {Le Borgne}, {Le Brun}, {Mignoli},
  {Pell{\'o}}, {Perez-Montero}, {Ricciardelli}, {Silverman}, {Tanaka},
  {Tresse}, {Bottini}, {Cappi}, {Cassata}, {Cimatti}, {Guzzo}, {Koekemoer},
  {Leauthaud}, {Maccagni}, {Marinoni}, {McCracken}, {Memeo}, {Meneux},
  {Porciani}, {Scaramella}, {Aussel}, {Capak}, {Halliday}, {Ilbert},
  {Kartaltepe}, {Salvato}, {Sanders}, {Scarlata}, {Scoville}, {Taniguchi}, \&
  {Thompson}}]{Bolzonella:2010aa}
{Bolzonella}, M., {Kova{\v c}}, K., {Pozzetti}, L., {et~al.} 2010, \aap, 524,
  A76

\bibitem[{{Borch} {et~al.}(2006){Borch}, {Meisenheimer}, {Bell}, {Rix}, {Wolf},
  {Dye}, {Kleinheinrich}, {Kovacs}, \& {Wisotzki}}]{borch06a}
{Borch}, A., {Meisenheimer}, K., {Bell}, E.~F., {et~al.} 2006, \aap, 453, 869

\bibitem[{{Bundy} {et~al.}(2006){Bundy}, {Ellis}, {Conselice}, {Taylor},
  {Cooper}, {Willmer}, {Weiner}, {Coil}, {Noeske}, \& {Eisenhardt}}]{bundy06a}
{Bundy}, K., {Ellis}, R.~S., {Conselice}, C.~J., {et~al.} 2006, \apj, 651, 120

\bibitem[{{Butcher} \& {Oemler}(1984)}]{Butcher:1984aa}
{Butcher}, H., \& {Oemler}, Jr., A. 1984, \apj, 285, 426

\bibitem[{{Chabrier}(2003)}]{Chabrier:2003aa}
{Chabrier}, G. 2003, \pasp, 115, 763

\bibitem[{{Coil} {et~al.}(2011){Coil}, {Blanton}, {Burles}, {Cool},
  {Eisenstein}, {Moustakas}, {Wong}, {Zhu}, {Aird}, {Bernstein}, {Bolton}, \&
  {Hogg}}]{Coil:2011aa}
{Coil}, A.~L., {Blanton}, M.~R., {Burles}, S.~M., {et~al.} 2011, \apj, 741, 8

\bibitem[{{Conroy} \& {Gunn}(2010)}]{Conroy:2010aa}
{Conroy}, C., \& {Gunn}, J.~E. 2010, {FSPS: Flexible Stellar Population
  Synthesis}, astrophysics Source Code Library, ascl:1010.043

\bibitem[{{Cool} {et~al.}(2013){Cool}, {Moustakas}, {Blanton}, {Burles},
  {Coil}, {Eisenstein}, {Wong}, {Zhu}, {Aird}, {Bernstein}, {Bolton}, {Hogg},
  \& {Mendez}}]{Cool:2013aa}
{Cool}, R.~J., {Moustakas}, J., {Blanton}, M.~R., {et~al.} 2013, \apj, 767, 118

\bibitem[{{Cooper} {et~al.}(2005){Cooper}, {Newman}, {Madgwick}, {Gerke},
  {Yan}, \& {Davis}}]{Cooper:2005aa}
{Cooper}, M.~C., {Newman}, J.~A., {Madgwick}, D.~S., {et~al.} 2005, \apj, 634,
  833

\bibitem[{{Cooper} {et~al.}(2008){Cooper}, {Newman}, {Weiner}, {Yan},
  {Willmer}, {Bundy}, {Coil}, {Conselice}, {Davis}, {Faber}, {Gerke},
  {Guhathakurta}, {Koo}, \& {Noeske}}]{cooper08a}
{Cooper}, M.~C., {Newman}, J.~A., {Weiner}, B.~J., {et~al.} 2008, \mnras, 383,
  1058

\bibitem[{{Cooper} {et~al.}(2010){Cooper}, {Coil}, {Gerke}, {Newman}, {Bundy},
  {Conselice}, {Croton}, {Davis}, {Faber}, {Guhathakurta}, {Koo}, {Lin},
  {Weiner}, {Willmer}, \& {Yan}}]{Cooper:2010aa}
{Cooper}, M.~C., {Coil}, A.~L., {Gerke}, B.~F., {et~al.} 2010, \mnras, 409, 337

\bibitem[{{Croton} {et~al.}(2006){Croton}, {Springel}, {White}, {De Lucia},
  {Frenk}, {Gao}, {Jenkins}, {Kauffmann}, {Navarro}, \&
  {Yoshida}}]{Croton:2006aa}
{Croton}, D.~J., {Springel}, V., {White}, S.~D.~M., {et~al.} 2006, \mnras, 365,
  11

\bibitem[{{Cucciati} {et~al.}(2010){Cucciati}, {Iovino}, {Kova{\v c}},
  {Scodeggio}, {Lilly}, {Bolzonella}, {Bardelli}, {Vergani}, {Tasca}, {Zucca},
  {Zamorani}, {Pozzetti}, {Knobel}, {Oesch}, {Lamareille}, {Caputi},
  {Kampczyk}, {Tresse}, {Maier}, {Carollo}, {Contini}, {Kneib}, {Le F{\`e}vre},
  {Mainieri}, {Renzini}, {Bongiorno}, {Coppa}, {de la Torre}, {de Ravel},
  {Franzetti}, {Garilli}, {Le Borgne}, {Le Brun}, {Mignoli}, {Pell{\`o}},
  {Peng}, {Perez-Montero}, {Ricciardelli}, {Silverman}, {Tanaka}, {Koekemoer},
  {Scoville}, {Abbas}, {Bottini}, {Cappi}, {Cassata}, {Cimatti}, {Guzzo},
  {Leauthaud}, {Maccagni}, {Marinoni}, {McCracken}, {Memeo}, {Meneux},
  {Porciani}, \& {Scaramella}}]{Cucciati:2010aa}
{Cucciati}, O., {Iovino}, A., {Kova{\v c}}, K., {et~al.} 2010, \aap, 524, A2

\bibitem[{{Dekel} \& {Birnboim}(2008)}]{Dekel:2008aa}
{Dekel}, A., \& {Birnboim}, Y. 2008, \mnras, 383, 119

\bibitem[{{Desai} {et~al.}(2007){Desai}, {Dalcanton}, {Arag{\'o}n-Salamanca},
  {Jablonka}, {Poggianti}, {Gogarten}, {Simard}, {Milvang-Jensen}, {Rudnick},
  {Zaritsky}, {Clowe}, {Halliday}, {Pell{\'o}}, {Saglia}, \&
  {White}}]{desai07a}
{Desai}, V., {Dalcanton}, J.~J., {Arag{\'o}n-Salamanca}, A., {et~al.} 2007,
  \apj, 660, 1151

\bibitem[{Dressler(1980)}]{dressler80a}
Dressler, A. 1980, \apj, 236, 351

\bibitem[{{Dressler}(1984)}]{dressler84a}
{Dressler}, A. 1984, \araa, 22, 185

\bibitem[{{Faber} {et~al.}(2007){Faber}, {Willmer}, {Wolf}, {Koo}, {Weiner},
  {Newman}, {Im}, {Coil}, {Conroy}, {Cooper}, {Davis}, {Finkbeiner}, {Gerke},
  {Gebhardt}, {Groth}, {Guhathakurta}, {Harker}, {Kaiser}, {Kassin},
  {Kleinheinrich}, {Konidaris}, {Kron}, {Lin}, {Luppino}, {Madgwick},
  {Meisenheimer}, {Noeske}, {Phillips}, {Sarajedini}, {Schiavon}, {Simard},
  {Szalay}, {Vogt}, \& {Yan}}]{Faber:2007aa}
{Faber}, S.~M., {Willmer}, C.~N.~A., {Wolf}, C., {et~al.} 2007, \apj, 665, 265

\bibitem[{{Fasano} {et~al.}(2000){Fasano}, {Poggianti}, {Couch}, {Bettoni},
  {Kj{\ae}rgaard}, \& {Moles}}]{Fasano:2000aa}
{Fasano}, G., {Poggianti}, B.~M., {Couch}, W.~J., {et~al.} 2000, \apj, 542, 673

\bibitem[{{Gallazzi} {et~al.}(2009){Gallazzi}, {Bell}, {Wolf}, {Gray},
  {Papovich}, {Barden}, {Peng}, {Meisenheimer}, {Heymans}, {van Kampen},
  {Gilmour}, {Balogh}, {McIntosh}, {Bacon}, {Barazza}, {B{\"o}hm}, {Caldwell},
  {H{\"a}ubler}, {Jahnke}, {Jogee}, {Lane}, {Robaina}, {Sanchez}, {Taylor},
  {Wisotzki}, \& {Zheng}}]{Gallazzi:2009aa}
{Gallazzi}, A., {Bell}, E.~F., {Wolf}, C., {et~al.} 2009, \apj, 690, 1883

\bibitem[{{Geha} {et~al.}(2012){Geha}, {Blanton}, {Yan}, \& {Tinker}}]{geha12a}
{Geha}, M., {Blanton}, M.~R., {Yan}, R., \& {Tinker}, J.~L. 2012, \apj, 757, 85

\bibitem[{{Gunn} \& {Gott}(1972)}]{Gunn:1972aa}
{Gunn}, J.~E., \& {Gott}, III, J.~R. 1972, \apj, 176, 1

\bibitem[{Guzzo {et~al.}(1997)Guzzo, Strauss, Fisher, Giovanelli, \&
  Haynes}]{guzzo97a}
Guzzo, L., Strauss, M.~A., Fisher, K.~B., Giovanelli, R., \& Haynes, M.~P.
  1997, \apj, 489, 37

\bibitem[{Hermit {et~al.}(1996)Hermit, Santiago, Lahav, Strauss, Davis,
  Dressler, \& Huchra}]{hermit96a}
Hermit, S., Santiago, B.~X., Lahav, O., {et~al.} 1996, \mnras, 283, 709

\bibitem[{{Hopkins} \& {Beacom}(2006)}]{hopkins06a}
{Hopkins}, A.~M., \& {Beacom}, J.~F. 2006, \apj, 651, 142

\bibitem[{Hubble(1936)}]{hubble36a}
Hubble, E.~P. 1936, The Realm of the Nebulae (New Haven: Yale University Press)

\bibitem[{{Hyde} \& {Bernardi}(2009)}]{Hyde:2009aa}
{Hyde}, J.~B., \& {Bernardi}, M. 2009, \mnras, 394, 1978

\bibitem[{{Iovino} {et~al.}(2010){Iovino}, {Cucciati}, {Scodeggio}, {Knobel},
  {Kova{\v c}}, {Lilly}, {Bolzonella}, {Tasca}, {Zamorani}, {Zucca}, {Caputi},
  {Pozzetti}, {Oesch}, {Lamareille}, {Halliday}, {Bardelli}, {Finoguenov},
  {Guzzo}, {Kampczyk}, {Maier}, {Tanaka}, {Vergani}, {Carollo}, {Contini},
  {Kneib}, {Le F{\`e}vre}, {Mainieri}, {Renzini}, {Bongiorno}, {Coppa}, {de la
  Torre}, {de Ravel}, {Franzetti}, {Garilli}, {Le Borgne}, {Le Brun},
  {Mignoli}, {Pell{\`o}}, {Peng}, {Perez-Montero}, {Ricciardelli}, {Silverman},
  {Tresse}, {Abbas}, {Bottini}, {Cappi}, {Cassata}, {Cimatti}, {Koekemoer},
  {Leauthaud}, {Maccagni}, {Marinoni}, {McCracken}, {Memeo}, {Meneux},
  {Porciani}, {Scaramella}, {Schiminovich}, \& {Scoville}}]{Iovino:2010aa}
{Iovino}, A., {Cucciati}, O., {Scodeggio}, M., {et~al.} 2010, \aap, 509, A40

\bibitem[{{Jarrett} {et~al.}(2000){Jarrett}, {Chester}, {Cutri}, {Schneider},
  {Skrutskie}, \& {Huchra}}]{Jarrett:2000aa}
{Jarrett}, T.~H., {Chester}, T., {Cutri}, R., {et~al.} 2000, \aj, 119, 2498

\bibitem[{{Johnston}(2011)}]{Johnston:2011aa}
{Johnston}, R. 2011, \aapr, 19, 41

\bibitem[{{Karim} {et~al.}(2011){Karim}, {Schinnerer},
  {Mart{\'{\i}}nez-Sansigre}, {Sargent}, {van der Wel}, {Rix}, {Ilbert},
  {Smol{\v c}i{\'c}}, {Carilli}, {Pannella}, {Koekemoer}, {Bell}, \&
  {Salvato}}]{Karim:2011aa}
{Karim}, A., {Schinnerer}, E., {Mart{\'{\i}}nez-Sansigre}, A., {et~al.} 2011,
  \apj, 730, 61

\bibitem[{{Kere{\v s}} {et~al.}(2005){Kere{\v s}}, {Katz}, {Weinberg}, \&
  {Dav{\'e}}}]{Keres:2005aa}
{Kere{\v s}}, D., {Katz}, N., {Weinberg}, D.~H., \& {Dav{\'e}}, R. 2005,
  \mnras, 363, 2

\bibitem[{{Kova{\v c}} {et~al.}(2010){Kova{\v c}}, {Lilly}, {Cucciati},
  {Porciani}, {Iovino}, {Zamorani}, {Oesch}, {Bolzonella}, {Knobel},
  {Finoguenov}, {Peng}, {Carollo}, {Pozzetti}, {Caputi}, {Silverman}, {Tasca},
  {Scodeggio}, {Vergani}, {Scoville}, {Capak}, {Contini}, {Kneib}, {Le
  F{\`e}vre}, {Mainieri}, {Renzini}, {Bardelli}, {Bongiorno}, {Coppa}, {de la
  Torre}, {de Ravel}, {Franzetti}, {Garilli}, {Guzzo}, {Kampczyk},
  {Lamareille}, {Le Borgne}, {Le Brun}, {Maier}, {Mignoli}, {Pello}, {Perez
  Montero}, {Ricciardelli}, {Tanaka}, {Tresse}, {Zucca}, {Abbas}, {Bottini},
  {Cappi}, {Cassata}, {Cimatti}, {Fumana}, {Koekemoer}, {Maccagni}, {Marinoni},
  {McCracken}, {Memeo}, {Meneux}, \& {Scaramella}}]{Kovac:2010ab}
{Kova{\v c}}, K., {Lilly}, S.~J., {Cucciati}, O., {et~al.} 2010, \apj, 708, 505

\bibitem[{{Kova{\v c}} {et~al.}(2014){Kova{\v c}}, {Lilly}, {Knobel},
  {Bschorr}, {Peng}, {Carollo}, {Contini}, {Kneib}, {Le F{\'e}vre}, {Mainieri},
  {Renzini}, {Scodeggio}, {Zamorani}, {Bardelli}, {Bolzonella}, {Bongiorno},
  {Caputi}, {Cucciati}, {de la Torre}, {de Ravel}, {Franzetti}, {Garilli},
  {Iovino}, {Kampczyk}, {Lamareille}, {Le Borgne}, {Le Brun}, {Maier},
  {Mignoli}, {Oesch}, {Pello}, {Montero}, {Presotto}, {Silverman}, {Tanaka},
  {Tasca}, {Tresse}, {Vergani}, {Zucca}, {Aussel}, {Koekemoer}, {Le Floc'h},
  {Moresco}, \& {Pozzetti}}]{Kovac:2014aa}
{Kova{\v c}}, K., {Lilly}, S.~J., {Knobel}, C., {et~al.} 2014, \mnras, 438, 717

\bibitem[{{Larson} {et~al.}(1980){Larson}, {Tinsley}, \&
  {Caldwell}}]{Larson:1980aa}
{Larson}, R.~B., {Tinsley}, B.~M., \& {Caldwell}, C.~N. 1980, \apj, 237, 692

\bibitem[{{Lauer} {et~al.}(2007){Lauer}, {Faber}, {Richstone}, {Gebhardt},
  {Tremaine}, {Postman}, {Dressler}, {Aller}, {Filippenko}, {Green}, {Ho},
  {Kormendy}, {Magorrian}, \& {Pinkney}}]{Lauer:2007aa}
{Lauer}, T.~R., {Faber}, S.~M., {Richstone}, D., {et~al.} 2007, \apj, 662, 808

\bibitem[{{Leja} {et~al.}(2013){Leja}, {van Dokkum}, \& {Franx}}]{Leja:2013aa}
{Leja}, J., {van Dokkum}, P., \& {Franx}, M. 2013, \apj, 766, 33

\bibitem[{{Lilly} {et~al.}(2009){Lilly}, {Le Brun}, {Maier}, {Mainieri},
  {Mignoli}, {Scodeggio}, {Zamorani}, {Carollo}, {Contini}, {Kneib}, {Le
  F{\`e}vre}, {Renzini}, {Bardelli}, {Bolzonella}, {Bongiorno}, {Caputi},
  {Coppa}, {Cucciati}, {de la Torre}, {de Ravel}, {Franzetti}, {Garilli},
  {Iovino}, {Kampczyk}, {Kovac}, {Knobel}, {Lamareille}, {Le Borgne}, {Pello},
  {Peng}, {P{\'e}rez-Montero}, {Ricciardelli}, {Silverman}, {Tanaka}, {Tasca},
  {Tresse}, {Vergani}, {Zucca}, {Ilbert}, {Salvato}, {Oesch}, {Abbas},
  {Bottini}, {Capak}, {Cappi}, {Cassata}, {Cimatti}, {Elvis}, {Fumana},
  {Guzzo}, {Hasinger}, {Koekemoer}, {Leauthaud}, {Maccagni}, {Marinoni},
  {McCracken}, {Memeo}, {Meneux}, {Porciani}, {Pozzetti}, {Sanders},
  {Scaramella}, {Scarlata}, {Scoville}, {Shopbell}, \&
  {Taniguchi}}]{Lilly:2009aa}
{Lilly}, S.~J., {Le Brun}, V., {Maier}, C., {et~al.} 2009, \apjs, 184, 218

\bibitem[{{Magnelli} {et~al.}(2009){Magnelli}, {Elbaz}, {Chary}, {Dickinson},
  {Le Borgne}, {Frayer}, \& {Willmer}}]{magnelli09a}
{Magnelli}, B., {Elbaz}, D., {Chary}, R.~R., {et~al.} 2009, \aap, 496, 57

\bibitem[{{Markwardt}(2009)}]{Markwardt:2009aa}
{Markwardt}, C.~B. 2009, in Astronomical Society of the Pacific Conference
  Series, Vol. 411, Astronomical Data Analysis Software and Systems XVIII, ed.
  D.~A. {Bohlender}, D.~{Durand}, \& P.~{Dowler}, 251

\bibitem[{{Martin} {et~al.}(2005){Martin}, {Fanson}, {Schiminovich},
  {Morrissey}, {Friedman}, {Barlow}, {Conrow}, {Grange}, {Jelinsky},
  {Milliard}, {Siegmund}, {Bianchi}, {Byun}, {Donas}, {Forster}, {Heckman},
  {Lee}, {Madore}, {Malina}, {Neff}, {Rich}, {Small}, {Surber}, {Szalay},
  {Welsh}, \& {Wyder}}]{Martin:2005aa}
{Martin}, D.~C., {Fanson}, J., {Schiminovich}, D., {et~al.} 2005, \apjl, 619,
  L1

\bibitem[{{Moore} {et~al.}(1998){Moore}, {Lake}, \& {Katz}}]{Moore:1998aa}
{Moore}, B., {Lake}, G., \& {Katz}, N. 1998, \apj, 495, 139

\bibitem[{{Morrissey} {et~al.}(2005){Morrissey}, {Schiminovich}, {Barlow},
  {Martin}, {Blakkolb}, {Conrow}, {Cooke}, {Erickson}, {Fanson}, {Friedman},
  {Grange}, {Jelinsky}, {Lee}, {Liu}, {Mazer}, {McLean}, {Milliard}, {Randall},
  {Schmitigal}, {Sen}, {Siegmund}, {Surber}, {Vaughan}, {Viton}, {Welsh},
  {Bianchi}, {Byun}, {Donas}, {Forster}, {Heckman}, {Lee}, {Madore}, {Malina},
  {Neff}, {Rich}, {Small}, {Szalay}, \& {Wyder}}]{Morrissey:2005aa}
{Morrissey}, P., {Schiminovich}, D., {Barlow}, T.~A., {et~al.} 2005, \apjl,
  619, L7

\bibitem[{{Moustakas} {et~al.}(2013){Moustakas}, {Coil}, {Aird}, {Blanton},
  {Cool}, {Eisenstein}, {Mendez}, {Wong}, {Zhu}, \&
  {Arnouts}}]{Moustakas:2013aa}
{Moustakas}, J., {Coil}, A.~L., {Aird}, J., {et~al.} 2013, \apj, 767, 50

\bibitem[{{Muldrew} {et~al.}(2012){Muldrew}, {Croton}, {Skibba}, {Pearce},
  {Ann}, {Baldry}, {Brough}, {Choi}, {Conselice}, {Cowan}, {Gallazzi}, {Gray},
  {Gr{\"u}tzbauch}, {Li}, {Park}, {Pilipenko}, {Podgorzec}, {Robotham},
  {Wilman}, {Yang}, {Zhang}, \& {Zibetti}}]{Muldrew:2012aa}
{Muldrew}, S.~I., {Croton}, D.~J., {Skibba}, R.~A., {et~al.} 2012, \mnras, 419,
  2670

\bibitem[{{Noeske} {et~al.}(2007){Noeske}, {Weiner}, {Faber}, {Papovich},
  {Koo}, {Somerville}, {Bundy}, {Conselice}, {Newman}, {Schiminovich}, {Le
  Floc'h}, {Coil}, {Rieke}, {Lotz}, {Primack}, {Barmby}, {Cooper}, {Davis},
  {Ellis}, {Fazio}, {Guhathakurta}, {Huang}, {Kassin}, {Martin}, {Phillips},
  {Rich}, {Small}, {Willmer}, \& {Wilson}}]{Noeske:2007aa}
{Noeske}, K.~G., {Weiner}, B.~J., {Faber}, S.~M., {et~al.} 2007, \apjl, 660,
  L43

\bibitem[{Norberg {et~al.}(2002)}]{norberg02a}
Norberg, P., {et~al.} 2002, \mnras, 332, 827

\bibitem[{Oemler(1974)}]{oemler74a}
Oemler, A. 1974, \apj, 194, 1

\bibitem[{{Peng} {et~al.}(2010){Peng}, {Lilly}, {Kova{\v c}}, {Bolzonella},
  {Pozzetti}, {Renzini}, {Zamorani}, {Ilbert}, {Knobel}, {Iovino}, {Maier},
  {Cucciati}, {Tasca}, {Carollo}, {Silverman}, {Kampczyk}, {de Ravel},
  {Sanders}, {Scoville}, {Contini}, {Mainieri}, {Scodeggio}, {Kneib}, {Le
  F{\`e}vre}, {Bardelli}, {Bongiorno}, {Caputi}, {Coppa}, {de la Torre},
  {Franzetti}, {Garilli}, {Lamareille}, {Le Borgne}, {Le Brun}, {Mignoli},
  {Perez Montero}, {Pello}, {Ricciardelli}, {Tanaka}, {Tresse}, {Vergani},
  {Welikala}, {Zucca}, {Oesch}, {Abbas}, {Barnes}, {Bordoloi}, {Bottini},
  {Cappi}, {Cassata}, {Cimatti}, {Fumana}, {Hasinger}, {Koekemoer},
  {Leauthaud}, {Maccagni}, {Marinoni}, {McCracken}, {Memeo}, {Meneux}, {Nair},
  {Porciani}, {Presotto}, \& {Scaramella}}]{Peng:2010aa}
{Peng}, Y.-j., {Lilly}, S.~J., {Kova{\v c}}, K., {et~al.} 2010, \apj, 721, 193

\bibitem[{{Pozzetti} {et~al.}(2010){Pozzetti}, {Bolzonella}, {Zucca},
  {Zamorani}, {Lilly}, {Renzini}, {Moresco}, {Mignoli}, {Cassata}, {Tasca},
  {Lamareille}, {Maier}, {Meneux}, {Halliday}, {Oesch}, {Vergani}, {Caputi},
  {Kova{\v c}}, {Cimatti}, {Cucciati}, {Iovino}, {Peng}, {Carollo}, {Contini},
  {Kneib}, {Le F{\'e}vre}, {Mainieri}, {Scodeggio}, {Bardelli}, {Bongiorno},
  {Coppa}, {de la Torre}, {de Ravel}, {Franzetti}, {Garilli}, {Kampczyk},
  {Knobel}, {Le Borgne}, {Le Brun}, {Pell{\`o}}, {Perez Montero},
  {Ricciardelli}, {Silverman}, {Tanaka}, {Tresse}, {Abbas}, {Bottini}, {Cappi},
  {Guzzo}, {Koekemoer}, {Leauthaud}, {Maccagni}, {Marinoni}, {McCracken},
  {Memeo}, {Porciani}, {Scaramella}, {Scarlata}, \&
  {Scoville}}]{Pozzetti:2010aa}
{Pozzetti}, L., {Bolzonella}, M., {Zucca}, E., {et~al.} 2010, \aap, 523, A13

\bibitem[{{Salim} {et~al.}(2007){Salim}, {Rich}, {Charlot}, {Brinchmann},
  {Johnson}, {Schiminovich}, {Seibert}, {Mallery}, {Heckman}, {Forster},
  {Friedman}, {Martin}, {Morrissey}, {Neff}, {Small}, {Wyder}, {Bianchi},
  {Donas}, {Lee}, {Madore}, {Milliard}, {Szalay}, {Welsh}, \&
  {Yi}}]{Salim:2007aa}
{Salim}, S., {Rich}, R.~M., {Charlot}, S., {et~al.} 2007, \apjs, 173, 267

\bibitem[{{Scodeggio} {et~al.}(2009){Scodeggio}, {Vergani}, {Cucciati},
  {Iovino}, {Franzetti}, {Garilli}, {Lamareille}, {Bolzonella}, {Pozzetti},
  {Abbas}, {Marinoni}, {Contini}, {Bottini}, {Le Brun}, {Le F{\`e}vre},
  {Maccagni}, {Scaramella}, {Tresse}, {Vettolani}, {Zanichelli}, {Adami},
  {Arnouts}, {Bardelli}, {Cappi}, {Charlot}, {Ciliegi}, {Foucaud}, {Gavignaud},
  {Guzzo}, {Ilbert}, {McCracken}, {Marano}, {Mazure}, {Meneux}, {Merighi},
  {Paltani}, {Pell{\`o}}, {Pollo}, {Radovich}, {Zamorani}, {Zucca}, {Bondi},
  {Bongiorno}, {Brinchmann}, {de La Torre}, {de Ravel}, {Gregorini}, {Memeo},
  {Perez-Montero}, {Mellier}, {Temporin}, \& {Walcher}}]{Scodeggio:2009aa}
{Scodeggio}, M., {Vergani}, D., {Cucciati}, O., {et~al.} 2009, \aap, 501, 21

\bibitem[{{Skibba} {et~al.}(2013){Skibba}, {Sheth}, {Croton}, {Muldrew},
  {Abbas}, {Pearce}, \& {Shattow}}]{Skibba:2013aa}
{Skibba}, R.~A., {Sheth}, R.~K., {Croton}, D.~J., {et~al.} 2013, \mnras, 429,
  458

\bibitem[{{Smith} {et~al.}(2005){Smith}, {Treu}, {Ellis}, {Moran}, \&
  {Dressler}}]{Smith:2005aa}
{Smith}, G.~P., {Treu}, T., {Ellis}, R.~S., {Moran}, S.~M., \& {Dressler}, A.
  2005, \apj, 620, 78

\bibitem[{{Swanson} {et~al.}(2008){Swanson}, {Tegmark}, {Hamilton}, \&
  {Hill}}]{Swanson:2008aa}
{Swanson}, M.~E.~C., {Tegmark}, M., {Hamilton}, A.~J.~S., \& {Hill}, J.~C.
  2008, \mnras, 387, 1391

\bibitem[{{Taylor} {et~al.}(2009){Taylor}, {Franx}, {van Dokkum}, {Bell},
  {Brammer}, {Rudnick}, {Wuyts}, {Gawiser}, {Lira}, {Urry}, \&
  {Rix}}]{taylor09a}
{Taylor}, E.~N., {Franx}, M., {van Dokkum}, P.~G., {et~al.} 2009, \apj, 694,
  1171

\bibitem[{{Tinker} {et~al.}(2011){Tinker}, {Wetzel}, \&
  {Conroy}}]{Tinker:2011aa}
{Tinker}, J., {Wetzel}, A., \& {Conroy}, C. 2011, ArXiv e-prints,
  arXiv:1107.5046

\bibitem[{{Weinmann} {et~al.}(2006){Weinmann}, {van den Bosch}, {Yang}, {Mo},
  {Croton}, \& {Moore}}]{Weinmann:2006aa}
{Weinmann}, S.~M., {van den Bosch}, F.~C., {Yang}, X., {et~al.} 2006, \mnras,
  372, 1161

\bibitem[{{West}(2005)}]{West:2005aa}
{West}, A.~A. 2005, PhD thesis, University of Washington, Washington, USA

\bibitem[{{West} {et~al.}(2010){West}, {Garcia-Appadoo}, {Dalcanton}, {Disney},
  {Rockosi}, {Ivezi{\'c}}, {Bentz}, \& {Brinkmann}}]{West:2010aa}
{West}, A.~A., {Garcia-Appadoo}, D.~A., {Dalcanton}, J.~J., {et~al.} 2010,
  \apj, 139, 315

\bibitem[{{Williams} {et~al.}(2009){Williams}, {Quadri}, {Franx}, {van Dokkum},
  \& {Labb{\'e}}}]{Williams:2009aa}
{Williams}, R.~J., {Quadri}, R.~F., {Franx}, M., {van Dokkum}, P., \&
  {Labb{\'e}}, I. 2009, \apj, 691, 1879

\bibitem[{{Wilman} {et~al.}(2010){Wilman}, {Zibetti}, \&
  {Budav{\'a}ri}}]{Wilman:2010aa}
{Wilman}, D.~J., {Zibetti}, S., \& {Budav{\'a}ri}, T. 2010, \mnras, 406, 1701

\bibitem[{{York} {et~al.}(2000){York}, {Adelman}, {Anderson}, {Anderson},
  {Annis}, {Bahcall}, {Bakken}, {Barkhouser}, {Bastian}, {Berman}, {Boroski},
  {Bracker}, {Briegel}, {Briggs}, {Brinkmann}, {Brunner}, {Burles}, {Carey},
  {Carr}, {Castander}, {Chen}, {Colestock}, {Connolly}, {Crocker}, {Csabai},
  {Czarapata}, {Davis}, {Doi}, {Dombeck}, {Eisenstein}, {Ellman}, {Elms},
  {Evans}, {Fan}, {Federwitz}, {Fiscelli}, {Friedman}, {Frieman}, {Fukugita},
  {Gillespie}, {Gunn}, {Gurbani}, {de Haas}, {Haldeman}, {Harris}, {Hayes},
  {Heckman}, {Hennessy}, {Hindsley}, {Holm}, {Holmgren}, {Huang}, {Hull},
  {Husby}, {Ichikawa}, {Ichikawa}, {Ivezi{\'c}}, {Kent}, {Kim}, {Kinney},
  {Klaene}, {Kleinman}, {Kleinman}, {Knapp}, {Korienek}, {Kron}, {Kunszt},
  {Lamb}, {Lee}, {Leger}, {Limmongkol}, {Lindenmeyer}, {Long}, {Loomis},
  {Loveday}, {Lucinio}, {Lupton}, {MacKinnon}, {Mannery}, {Mantsch}, {Margon},
  {McGehee}, {McKay}, {Meiksin}, {Merelli}, {Monet}, {Munn}, {Narayanan},
  {Nash}, {Neilsen}, {Neswold}, {Newberg}, {Nichol}, {Nicinski}, {Nonino},
  {Okada}, {Okamura}, {Ostriker}, {Owen}, {Pauls}, {Peoples}, {Peterson},
  {Petravick}, {Pier}, {Pope}, {Pordes}, {Prosapio}, {Rechenmacher}, {Quinn},
  {Richards}, {Richmond}, {Rivetta}, {Rockosi}, {Ruthmansdorfer}, {Sandford},
  {Schlegel}, {Schneider}, {Sekiguchi}, {Sergey}, {Shimasaku}, {Siegmund},
  {Smee}, {Smith}, {Snedden}, {Stone}, {Stoughton}, {Strauss}, {Stubbs},
  {SubbaRao}, {Szalay}, {Szapudi}, {Szokoly}, {Thakar}, {Tremonti}, {Tucker},
  {Uomoto}, {Vanden Berk}, {Vogeley}, {Waddell}, {Wang}, {Watanabe},
  {Weinberg}, {Yanny}, {Yasuda}, \& {SDSS Collaboration}}]{York:2000aa}
{York}, D.~G., {Adelman}, J., {Anderson}, Jr., J.~E., {et~al.} 2000, \aj, 120,
  1579

\bibitem[{Zehavi {et~al.}(2002)}]{zehavi02a}
Zehavi, I., {et~al.} 2002, \apj, 571, 172

\end{thebibliography}

\end{document}